-------------- Enclosure number 1 ----------------
\documentclass{iopart}
\usepackage{amssymb}

\def\bbbc{{\Bbb C}}
\def\bbbe{{\Bbb E}}

\newtheorem{remark}{Remark}
\newtheorem{example}{Example}
\def\re{\mbox{Re}\,}
\def\im{\mbox{Im}\,}

\def\fr{\frak }

\begin{document}
\jl{1}
\def\bbbe{I\!\!\!\!E}
\begin{flushright}
{\bf solv-int/9909020}
\end{flushright}

\title[The complex Toda chains and the simple Lie algebras]{The complex
Toda chains and the simple Lie \\[5pt] algebras -- solutions and large
time asymptotics  -- II} \bigskip

\author{V S Gerdjikov\dag, E G Evstatiev\dag\ddag\ and R I
Ivanov\dag$^\ast$}

\address{\dag\ Institute for Nuclear Research and Nuclear Energy,
Bulg. Acad. of Sci.,\\ boul. Tzarigradsko shosse 72,
1784 Sofia,Bulgaria }

\address{\ddag\ Department of Physics, University of Texas at Austin,
\\ Austin, Texas 78712, USA}

\address{$^\ast $ Department of Mathematical Physics \\
National University of Ireland - Galway, Galway, Ireland}

\begin{abstract}
We propose a compact and explicit expression for the solutions of the
complex Toda chains related to the classical series of simple Lie
algebras ${\frak  g} $. The solutions are parametrized by a minimal set of
scattering data for the corresponding Lax matrix. They are expressed as
sums over the weight systems of the fundamental representations of
${\frak g} $ and are explicitly covariant under the corresponding Weyl
group action. In deriving these results we start from the Moser formula
for the ${\bf A}_r $ series and obtain the results for the other classical
series of Lie algebras by imposing appropriate involutions on the
scattering data. Thus we also show how Moser's solution goes into the one
of Olshanetsky and Perelomov. The results
for the large-time asymptotics of the ${\bf A}_r $--CTC solutions are
extended to the other classical series ${\bf B}_r $--${\bf D}_r $. We
exhibit also some `irregular' solutions for the ${\bf D}_{2n+1} $ algebras
whose asymptotic regimes at $t\to\pm\infty  $ are qualitatively different.
Interesting examples of bounded and periodic solutions are presented and
the relations between the solutions for the algebras ${\bf D}_4 $, ${\bf
B}_3 $ and ${\bf G}_2 $ are analyzed.

\end{abstract}
\vfill

\centerline{Sofia, September, 1999}
\eqnobysec

\section{Introduction}\label{sec:in}

The famous Toda chain model
\cite{Toda1,Man,Fla,Moser,Bogoy1,OlshPer,LeSav1,Mih1,MihOlPer1,OlshPer1,%
LeSav,Toda} was initially introduced in order to study the
nearest neighbor interactions in atomic chains. Soon it was shown that it
also possesses interesting mathematical properties and that to each simple
Lie algebra ${\frak  g}$ one can relate a natural generalization to the
Toda chain \cite{LeSav,Bogoy1,OlshPer,MihOlPer1,OlshPer1,LeSav1,%
Kosta,Symes,Goodman,OPRS},
namely:
\begin{equation}\label{eq:gCTC}
{d^2 \vec{q}  \over dt^2 }
= \sum_{k=1}^{r} \alpha_k  \rme^{-(\vec{q},\alpha_k) },
\end{equation}
where $\vec{q}=(q_1,\dots,q_r) $ is a vector in the root space $\bbbe^r $
of the algebra ${\frak  g} $ of rank $r $ and $\alpha _k $, $k=1,2,\dots,r
$ are the simple roots of ${\frak  g} $. One may view $q_k(t) $ as the
coordinate of the $k $-th particle and study the effect of their
interaction. A number of results in this direction are known showing how
the (real) Toda chain (RTC) (\ref{eq:gCTC}) can be viewed as completely
integrable Hamiltonian system and how it can be explicitly solved, see
\cite{Man,Fla,Moser,Bogoy1,OlshPer,LeSav1,MihOlPer1,OlshPer1,LeSav,%
Kosta,Symes,OPRS}.

Recently it became known that generalizing the RTC model with ${\frak
g}\simeq sl(N)$ to `complex' particles (i.e., now $q_k(t) $ become
complex-valued functions) allows one to describe the interactions in the
$N $-soliton trains of the nonlinear Schr\"odinger equation in the
adiabatic approximation. In this case each soliton behaves as separate
entity (`particle'); $\re q_k(t) $ describes its center of mass position
and $\im q_k(t)$ determines its phase, for more details see
\cite{GKUE,GUED,Arnold,GEI}. These facts draw our interest towards the
study of the complex Toda chain (CTC) models when the dynamical variables
$q_k(t) $ become complex, while the time variable $t $ stays real.

It is well known that a large number of results for the RTC are trivially
generalized to the CTC case by just making the corresponding parameters
complex. These include the Lax pairs and the explicit solutions. However,
since each `complex' particle has two degrees of freedom their interaction
becomes much more complex and qualitatively different as compared to the
real case. In particular the set of asymptotic regimes for the CTC is much
richer than the ones of RTC. In addition to the asymptotically free
particle regime (the only one possible for RTC), CTC allows also for bound
state regimes, mixed regimes, degenerate regimes, etc. These facts were
reported in \cite{GKUE,GUED,Arnold,GEI}.

The present paper is a natural extension of \cite{GEI}; it also contains
the proofs and generalizations of the results in \cite{GEI}.

There are several methods for solving the RTC which readily generalize
also for CTC. The method in \cite{OlshPer,OPRS,Kosta,Symes}
allows one to write down the solution as:
\begin{equation}\label{eq:q-ome}
(\vec{q}(t),\omega _k) - (\vec{q}(0),\omega _k) =
\ln \langle \omega_k | \rme^{-2L(0) t} |\omega_k \rangle,
\end{equation}
where $\omega _k $ is the $k $-th fundamental weight of ${\fr g} $,
$|\omega _k\rangle  $ is the highest weight vector in the $k $-th
fundamental representation $R(\omega _k) $ and $L(0) $ is the Lax matrix
evaluated at $t=0 $, see formula (\ref{eq:g-laxa}) below. The right hand
side of (\ref{eq:q-ome}) has the obvious advantage of being written in
compact and invariant form. However it is difficult to extract from it
explicitly parametrized solutions.

Another well known approach to solving the Toda chain models was developed
in \cite{LeSav}. It allows one to express the solution in terms of $2r $
constants. Starting from a comparatively simple expression for
$X_1=\exp(-q_1(t)) $ one then calculates $X_k=\exp(-q_k(t)) $  as a
determinant of a $k\times k $ matrix whose elements are determined by the
derivatives of $X_1$, see \cite{LeSav1,LeSav}. One may also use a
recurrent procedure to evaluate $ X_k $. However this leads to rather
complicated and difficult to analyze expressions.

Our first aim in the present paper will be to analyze (\ref{eq:q-ome}) and
write it down in the  form:
\numparts\label{eq:**}
\begin{eqnarray}\label{eq:**a}
(\vec{q}(t),\omega _k) = \ln {\cal  B}_{{\frak  g};k}(t), \\
\label{eq:**b}
{\cal  B}_{{\frak  g};k}(t)= \sum_{\gamma \in \Gamma_{{\frak  g}} (\omega
_k)} \exp\left[ -2(\gamma,\vec{\zeta })t + (\vec{\varphi }_0,\gamma
)\right] W^{(k)}(\vec{\zeta },\gamma ).
\end{eqnarray}
\endnumparts
Here $(\gamma,\vec{\zeta }) $ is the scalar product between the vector
$\vec{\zeta }=(\zeta _1,\zeta _2,\dots,\zeta _r) $ and the weight $\gamma
\in \Gamma_{\frak g} (\omega _k) $; $\zeta _s $ are eigenvalues of $L(0) $
and we suppose that they satisfy $\zeta _k\neq \zeta _j $ for $k\neq j $.
The components of the vectors $\vec{\zeta } $ and $\vec{\varphi}_0
=(\varphi _{01}, \varphi _{02}, \dots, \varphi _{0r})$ provide the $2r $
(complex) parameters directly related to the minimal set of scattering
data ${\cal T}_{{\frak  g}}$ of $L(0) $, see formula (\ref{eq:Tau-g})
below; $\Gamma_{{\frak  g}} (\omega _k) $ is the set of weights of the $k
$-th fundamental representation of ${\frak g} $; $W^{(k)}(\vec{\zeta},
\gamma ) $ are  $t $-independent functions which are defined in
Section~3 below.
Thus the right hand side of (\ref{eq:**a}), (\ref{eq:**b}) like the right
hand side of (\ref{eq:q-ome}) is invariant and at the same time is
explicitly parametrized. This fact allows us to calculate explicitly the
large time asymptotics of $\vec{q}(t) $:
\begin{equation}\label{eq:***}
\lim_{t\to\pm\infty } (\vec{q}(t) - \vec{v}^\pm t) =\vec{\varphi }_0^\pm +
\vec{\beta }^\pm.
\end{equation}
It is a well known fact \cite{Goodman,OPRS} that $\vec{v}^+=
w_0(\vec{v}^-)= -2\vec{\zeta } $, where $w_0 $ is the Weyl
group element which maps the highest weight $\omega _k $ of the $k $-th
fundamental representation $R(\omega _k)$ of ${\frak  g} $ into the
corresponding lowest weight vector $\omega _k^- $.  We provide explicit
expressions for $\vec{\beta }^\pm $ as functions of $\vec{\zeta } $ and
also show that
\begin{equation}\label{eq:beta-w}
\vec{\beta }^+(\vec{\zeta } )=\vec{\beta }^-(w_0(\vec{\zeta }) ), \qquad
\vec{\varphi }_0^+ = w_0(\vec{\varphi }_0^-) = \vec{\varphi }_0.
\end{equation}

In fact the solutions to the CTC related to a certain simple Lie algebra
${\frak  g} $ may be derived in two ways. The first one is to cast the
solution (\ref{eq:q-ome}) in the form
\begin{equation}\label{eq:q-m}
\langle \omega_k | \rme^{-2L(0) t} |\omega_k \rangle =
\sum_{\gamma \in \Gamma_{{\frak  g}} (\omega_k)}^{}  (\langle \omega _k
|V| \gamma \rangle )^2 \rme^{-2(\vec{\zeta },\gamma )t}
\end{equation}
and then try to evaluate the matrix elements $\langle \omega _k |V| \gamma
\rangle $ in terms of the scattering data ${\cal  T}_{{\frak  g}} $. This
requires the explicit construction of $V $ for each of the fundamental
representations $R(\omega _k) $ of ${\frak  g} $.

The second possibility which will be used below, is to start with the well
known solution of Moser \cite{Moser} for $sl(N) $ with conveniently
chosen $N $ and impose on the scattering data the involution that
restricts it to ${\frak g} $.  Obviously both solutions must coincide. The
proof of this fact is also one of the results in the present paper.

In the next section we introduce the notations and analyze the properties
of the fundamental representations of the classical series of simple Lie
algebras and derive some useful relations between the matrix elements of
the typical and the other fundamental representations.  In Section~3 we
prove formula (\ref{eq:**a}), (\ref{eq:**b}) for each of the classical
series ${\bf A}_r $ -- ${\bf D}_r $. In Section~4 we extend the results
for the large-time asymptotics  of the ${\bf A}_r $--CTC solutions also to
the other classical series ${\bf B}_r $--${\bf D}_r $. We exhibit also
some `irregular' solutions for the ${\bf D}_{2n+1} $ algebras whose
asymptotic regimes at $t\to\pm\infty $ are qualitatively different.  We
provide also some interesting examples of bounded and periodic solutions
and analyze the relations between the solutions for the algebras ${\bf
D}_4 $, ${\bf B}_3 $ and ${\bf G}_2 $.

\section{Preliminaries}

In what follows we shall use the so-called `symmetric' Lax representation
for the CTC model (\ref{eq:gCTC}):
\numparts\label{eq:g-lax}
\begin{equation}\label{eq:g-laxa}
L(t)=\sum_{k=1}^{r} \left( b_k H_k + a_k(E_{\alpha_k }+
E_{-\alpha_k }) \right),
\end{equation}
\begin{equation}\label{eq:g-laxb}
M(t)= \sum_{k=1}^r a_k(E_{\alpha_k } - E_{-\alpha_k }),
\end{equation}
\endnumparts
where $a_k={1 \over 2}\rme^{-(q,\alpha_k)/2} $ and $b_k={1\over 2}
{dq_k/dt } $. For ${\fr g}\simeq sl(N) $ we have $a_k={1 \over
2}\rme^{(q_{k+1}-q_k)/2} $.  It is well known that to each root $\alpha$
from the root system $\Delta _g\subset \bbbe^r $ one can relate the
element $H_\alpha $ of the Cartan subalgebra $ {\frak h}\subset {\frak  g}
$.  Analogously, to $q(t) = \mbox{Re\,} q(t) + i \,\mbox{Im\,} q(t) \in
{\frak  h}$ there corresponds the vector $\vec{q}(t) = \mbox{Re\,}
\vec{q}(t)+ i\, \mbox{Im\,} \vec{q}(t) $, whose real and imaginary parts
are vectors in the root space $\bbbe^r $.

The integrals of motion in involution for the CTC model are provided by
the eigenvalues, $\zeta_k=\kappa _k+i\eta_k $, of $L $. The  solutions of
both the CTC and the RTC are determined by the scattering data for $L(0)
$. When the spectrum of $L(0) $ is nondegenerate, i.e.  $\zeta _k \neq
\zeta _j $ for $k\neq j $, then this scattering data consists of
\begin{equation}\label{eq:Tau}
{\cal  T}\equiv \{ \zeta _1, \dots, \zeta _N, r_1, \dots, r_N\}.
\end{equation}
where $r_k $ are the first components of the corresponding eigenvectors
$v^{(k)} $ of $L(0) $ in the typical representation $R(\omega _1) $
of $\fr g $, $N=\mbox{dim\,}R(\omega_1) $. If we combine all eigenvectors
$v^{(k)} $ as columns of the matrix $V $ then $r_{k} = V_{1k} $ and
\begin{equation}\label{eq:eigve}
L(0) V= V Z, \qquad Z=\mbox{diag\,}(\zeta _1, \dots, \zeta_N).
\end{equation}
It is known that the eigenvectors of symmetric matrices $L(0) $  with
nondegenerate spectrum can always be normalized, i.e. following
\cite{Moser,LeSav,Toda} we require that
\begin{equation}\label{eq:norm}
 (v^{(k)}, v^{(k)}) \equiv \sum_{s=1}^{N} \left( V_{sk}\right)^2 = 1,
\qquad k=1,\dots, N;
\end{equation}
and besides $V^T=V^{-1} $. Equation (\ref{eq:norm}) determines $r_k $
up to a sign.

{}From (\ref{eq:eigve}) it follows  that
\begin{eqnarray}\label{eq:1-1}
G(t)\equiv \rme^{-2L(0)t} = V \rme^{-2Zt} V^{-1},
\end{eqnarray}
and we can rewrite (\ref{eq:q-ome}) in the form (\ref{eq:**a}),
(\ref{eq:**b}) with
\begin{eqnarray}\label{eq:1-1c}
{\cal  B}_{{\frak  g};k}(t)= \sum_{\gamma \in \Gamma_{{\frak  g}} (\omega
_k)}^{} (\langle \omega _k |V| \gamma \rangle )^2 \rme^{-2(\vec{\zeta
},\gamma )t}
\end{eqnarray}
Thus our aim will be realized if we obtain explicit expressions for the
matrix elements $\langle \omega _k | V|\gamma \rangle $ of $V $ in the $k
$-th fundamental representation in terms of ${\cal  T} $ which is
determined by the spectral data of $L(0) $ in the typical representation
$R(\omega _1) $ of $\frak g$.

The eigenvalues of $L(0)$, and especially their real parts $\kappa _k $,
which can be calculated directly from the initial conditions as it will
become evident below, uniquely determine the asymptotic behavior of the
solutions \cite{GEI}.  We will extensively use this fact for the
description of the different types of asymptotic behavior.

The minimal set of scattering data for ${\fr g}\sim sl(N) $  is obtained
from (\ref{eq:Tau}) by imposing on ${\cal  T} $ the restrictions
$\sum_{k=1}^{N}\zeta _k =0 $ and
\begin{equation}\label{eq:norm1}
\sum_{k=1}^{N} r_k^2 =1,
\end{equation}
which follows from $V^T=V^{-1} $. Therefore one may consider as
${\cal  T}_{{\bf A}_r} $ the set
\begin{eqnarray}\label{eq:tau-Ar}
{\cal  T}_{A_r} &\equiv& \left\{ \zeta _1,\dots,\zeta _N; \tilde{\varphi }
_{01}, \dots, \tilde{\varphi} _{0N} \right\}, \qquad N=r+1,\nonumber\\
\tilde{\varphi} _{0k} &=& \ln r_k^2 - {1  \over N } \sum_{s=1}^{N} \ln
r_s^2.
\end{eqnarray}
Although the number of elements in ${\cal  T}_{{\bf A}_r} $ is $2N $
(instead of $2r=2N-2 $) it is obvious that only $2r $ of them are
independent.

{}For the other classical series of simple Lie algebras the elements of
${\cal  T} $ (\ref{eq:Tau}) satisfy symmetry relations, namely \cite{GEI}:
\begin{eqnarray}\label{eq:26a}
\zeta _k &=& -\zeta _{\bar{k}}, \qquad \bar{k} = N+1-k,\\
\label{eq:26b}
r_k r_{\bar{k}} &=& \rme^{-q_1(0)} w_k,
\end{eqnarray}
for $k=1,\dots,N $ where $N $ is the dimension of the typical
representation $R(\omega _1) $ and the value of $q_1(t) $ at $t=0 $ is
determined through the normalization condition (\ref{eq:norm1}). The
coefficients $w_k $ are time independent and are expressed in terms of
$\zeta_1, \dots, \zeta _r $ as follows, see Appendix~A.

{\bf ${\bf B}_r $-series: $N=2r+1 $.} Note that in this case $\zeta
_{r+1}=0 $,
\begin{equation}\label{eq:w-k-b}
w_k = {1  \over  8\zeta _k^{2}} \prod_{s=1}^{k-1} {1  \over  4\zeta_s^2
- 4\zeta_k^2} \prod_{s=k+1}^{r} {1  \over 4\zeta _k^2 - 4\zeta _s^2},
\end{equation}
and in addition to (\ref{eq:26b}):
\begin{equation}\label{eq:r0-b}
r_{r+1}^2 =  \rme^{-q_1(0)} w_{r+1}  \qquad w_{r+1}  =
\prod_{s=1}^{r} {1  \over 4\zeta _s^2 }.
\end{equation}
Inserting (\ref{eq:26b})--(\ref{eq:r0-b}) into (\ref{eq:norm1}) we
obtain a quadratic equation for $\exp{(-q_1(0))} $, so it can be expressed
in terms of ${\cal T}_{g}$.

{\bf ${\bf C}_r $-series: $N=2r $} and
\begin{equation}\label{eq:w-k-c}
w_k = -{1  \over  4\zeta _k} \prod_{s=1}^{k-1} {1  \over  4\zeta _s^2
- 4\zeta_k^2} \prod_{s=k+1}^{r} {1  \over 4\zeta _k^2 - 4\zeta _s^2},
\end{equation}

{\bf ${\bf D}_r $-series: $N=2r $} and
\begin{equation}\label{eq:w-k-d}
w_k = \prod_{s=1}^{k-1} {1  \over 4\zeta _s^2
- 4\zeta_k^2} \prod_{s=k+1}^{r} {1  \over 4\zeta _k^2 - 4\zeta _s^2},
\end{equation}
In the last two cases again $\exp{(-q_1(0))} $ is determined from
(\ref{eq:norm1}).  The derivation of the solution for  the ${\bf D}_r $
series requires some additional efforts. The main problem here is related
with the treatment of the spinor representations.

The proof of (\ref{eq:w-k-b})--(\ref{eq:w-k-d}) is based on the study of
the properties of the corresponding matrices $V $ and is given in the
Appendix~A. Then one easily finds that the set of parameters
\begin{equation}\label{eq:Tau-g}
{\cal  T}_{{\frak  g}} \equiv \left\{ \zeta _1,\dots, \zeta _r; \varphi
_{01}, \dots, \varphi _{0r} \right\}, \qquad \varphi _{0k}= \ln (r_k /
r_{\bar{k}}),
\end{equation}
uniquely determines ${\cal  T}$ (\ref{eq:Tau}), which in turn provides the
full set of eigenvalues and eigenvectors of $L(0) $.

Next we will need a number of details from the representation theory of
the simple Lie algebras. In what follows by $R_{\frak g}(\omega ) $ we
will denote the representation of ${\frak  g} $ with highest weight
$\omega  $; $\Gamma_{{\frak  g}} (\omega ) $ stands for the set of weights
of $R_{{\frak  g}} (\omega )$.  Often when the choice for ${\frak  g} $
is clear from the context, we will omit the subscript and will write
simply $\Gamma (\omega ) $ and $ R(\omega ) $. We will also need to
introduce ordering not only in the root system $\Delta _{{\frak g}} $ but
also in the weight system $R_{{\frak g}} (\omega )$. To this end we will
use a vector $\vec{K} $ in the root space $\bbbe^r $ such that $(\gamma _1
- \gamma _2, \vec{K}) \neq 0$ and $(\alpha  - \beta, \vec{K}) \neq 0$ for
any two weights $\gamma _1 \neq \gamma _2 \in \Gamma _{{\frak g}}(\omega )
$ and roots $\alpha \neq \beta \in \Delta _{{\frak g}} $.  Without
restrictions we can choose $\vec{K} $, together with the vector
$-\vec{\kappa }=-\re \vec{\zeta } $ to be in the fundamental Weyl chamber
so that $(\omega -\gamma, \vec{K})>0$  for any $\gamma \in \Gamma _{{\frak
g}} (\omega ) $.

Let us now denote by $\gamma _k $, $k=1,\dots, N $ the set of weights of
the typical representation $\Gamma _{{\frak  g}}(\omega _1) $ of ${\frak
g} $, namely:
\numparts\label{eq:Al-1}
\begin{equation}\label{eq:Al-1a}
\gamma _k = e_k - {1  \over r+1 } \sum_{a=1}^{r+1} e_a, \qquad N=r+1,
\end{equation}
for ${\frak  g}\simeq {\bf A}_r$,
\begin{equation}\label{eq:Al-1cd}
\gamma _k = \cases{e_k &for  $1\leq k \leq r$\\
-e_{\bar{k}} &for  $r+1 \leq k \leq 2r$\\}  \qquad N=2r,
\end{equation}
for ${\frak  g}\simeq {\bf C}_r, {\bf D}_r$,
\begin{equation}\label{eq:Al-1b}
\gamma _k = \cases{ e_k & for  $1\leq k \leq r $\\
0 &for $ k=r+1$ \\
-e_{\bar{k}} & for $ r+2 \leq k \leq 2r+1$\\} \qquad
N=2r+1,
\end{equation}
\endnumparts
for ${\frak  g}\simeq {\bf B}_r$; in all formulae above
$\bar{k} = N+1-k $. The corresponding weight vectors specify an
orthonormal basis in $R_{{\frak  g}}(\omega _1) $ and will be denoted by
$|\gamma _k\rangle  $.

An important and well known tool to construct the fundamental
representations of ${\frak  g} $ is to make use of the exterior tensor
products of $R_{{\frak  g}}(\omega _1) $. Indeed, the orthonormal basis in
$\wedge^k R_{{\frak  g}}(\omega _1) $ consists of the weight vectors
\begin{eqnarray}\label{eq:Al-2}
\fl |\gamma _{I}\rangle \equiv |\gamma _{i_1,i_2,\dots,i_k}\rangle =
|\gamma _{i_1}\wedge \gamma _{i_2}\wedge \dots \wedge \gamma
_{i_k}\rangle, \qquad I\equiv \{i_1,i_2,\dots,i_k\},
\end{eqnarray}
with $1\leq i_1 <i_2<\dots <i_k\leq N $; the weight corresponding to
(\ref{eq:Al-2}) is obviously:
\begin{eqnarray}\label{eq:Al-3}
\gamma _{I}= \gamma _{i_1} + \gamma _{i_2} + \dots +
\gamma _{i_k}.
\end{eqnarray}
Here and in what follows we shall use the one-to-one correspondence
between the set of indices $I $ and the corresponding weight $\gamma_{I} $
and weight vector $|\gamma _{I} \rangle $.

{}For ${\frak  g}\simeq {\bf A}_r $ all fundamental representations
are in fact exterior powers of the typical one $R(\omega _1) $:
\begin{eqnarray}\label{eq:Al-4}
R(\omega _k)= \wedge^k R(\omega _1),
\end{eqnarray}
for $k=1,2,\dots,r=N-1 $ and $\gamma _I=\gamma _I^{(k)} $ where the upper
index $k $ means that $\gamma _I^{(k)}\in \Gamma (\omega _k) $.  We remind
also other well known fact, namely that in $\wedge^k R(\omega _1) $ we
have
\begin{eqnarray}\label{eq:Al-5}
\fl\langle \omega _k |V |\gamma _{I}^{(k)}\rangle \equiv
\langle \gamma _1 \wedge \gamma _2 \wedge \dots \wedge \gamma _k | V |
\gamma _{i_1}\wedge \gamma _{i_2} \wedge \dots \wedge \gamma _{i_k}
\rangle =  V\left\{ \begin{array}{cccc} 1 & 2 & \dots & k \\ i_1 & i_2 &
\dots &i_k \end{array} \right\},
\end{eqnarray}
where $V\left\{ \begin{array}{cccc} j_1 & j_2 & \dots & j_k \\ i_1 &
i_2 & \dots &i_k \end{array} \right\} $ is the minor  of the group
element $V\in {\frak  G} $ determined by the intersection of the rows $j_1,
j_2,\dots, j_k$ with the columns $i_1,i_2,\dots, i_k $. Thus given a
group element $V\in SL(r+1) $ in the typical representation $R(\omega _1)
$ one can construct its image for each of the fundamental representations
$R(\omega _k) $, $k=1,2,\dots,r $.

Let us now explain how this can be done for the other simple Lie algebras
of the classical series. To this end we shall make use of the well known
facts about the root systems \cite{Bour,GotGro*78} of $\frak g $ and about
the tensor products of their fundamental representations, see
\cite{VinOni}

Let us now consider the ${\bf B}_r $ series. Then we have:
\begin{eqnarray}\label{eq:Al-6}
\wedge^k R(\omega _1) &=& R(\omega _k), \qquad \mbox{for} \;
k=1,2,\dots, r-1, \nonumber\\
\wedge^r R(\omega _1) &=& R(2\omega _r),
\end{eqnarray}
where
\begin{equation}\label{eq:Al-6om}
\omega _k= e_1+e_2+\dots + e_k, \qquad
\omega _r = \case{1}{2} (e_1 + e_2 +\dots +e_r)
\end{equation}
Here $\omega _r $ is the highest weight of the spinor representation of
${\bf B}_r $. Therefore the relations (\ref{eq:Al-5}) hold also for
${\frak  g}\simeq {\bf B}_r $ with $k=1,2,\dots,r-1 $.

Other well known fact is that the symmetric tensor product of $R(\omega
_r) $ is generically reducible and
\begin{equation}\label{eq:Al-9}
R(\omega _r) \mathop{\otimes}\limits_S R(\omega _r) = R(2\omega _r)
\oplus \sum_{i=1}^{[ r/4]} R(\omega _{r-4i+1}) \oplus R(\omega _{r-4i}),
\end{equation}
where $\omega _0=0 $ and $R(\omega _0) $ is the trivial $1 $-dimensional
representation of $ {\frak  g} $.

We have now two possibilities to introduce a basis in $R(2\omega _r) $. The
first one is to use (\ref{eq:Al-5}) and (\ref{eq:Al-6}) like above. The
second possibility is to argue that $R(\omega _r)
\mathop{\otimes}\limits_S R(\omega _r)  $ is spanned by:
\begin{equation}\label{eq:R-2om-r}
|\gamma _{I}^{(r)} \otimes \gamma _{I}^{(r)}\rangle  \qquad \mbox{and}
\qquad {1  \over \sqrt{2} } \left( |\gamma _{I}^{(r)} \otimes \gamma
_{J}^{(r)}\rangle  + |\gamma _{J}^{(r)} \otimes \gamma _{I}^{(r)}\rangle
\right),
\end{equation}
where $\gamma _{I}^{(r)}\in \Gamma (\omega _r) $ are weights of the spinor
representation of ${\bf B}_r $. Obviously they have the form:
\begin{equation}\label{eq:Al-166}
\gamma _{I}^{(r)}={1  \over 2 } \sum_{k=1}^{r} \sigma _ke_k =
{1  \over 2 }\sum_{k=1}^{r} \gamma _{i_k},
\end{equation}
where $\sigma _k=\pm 1 $. The corresponding sets $I $ (and $J $) now must
be special in the sense that: a)~$r+1\not\in I $; b)~if $i_k\in I $ then
$\bar{i}_k = N+1-i_k \not\in I $. In other words $I $ does not contain
pairs of `conjugated` indices. To the weight vector  $|\gamma _{I}^{(r)}
\otimes \gamma _{I}^{(r)}\rangle  $ there corresponds the weight $2\gamma
_{I}^{(r)} $ which can be obtained from $2\omega _r $ by a Weyl group
transformation. This means that $|\gamma _{I}^{(r)} \otimes  \gamma
_{I}^{(r)}\rangle  $ belongs to the module $R(2\omega _r) $ in the right
hand side of (\ref{eq:Al-9}). Thus we find:
\begin{eqnarray}\label{eq:Al-10}
V\left\{ \begin{array}{cccc} 1 & 2 & \dots & r\\ i_1 & i_2 & \dots & i_r
\end{array} \right\} =\langle 2\omega _r|V|2\gamma_{I}^{(r)}\rangle
\nonumber\\
= \langle \omega _r\otimes \omega _r |V|\gamma_{I}^{(r)}\otimes
\gamma_{I}^{(r)} \rangle= \left(\langle \omega _r |V|\gamma_{I}^{(r)}
\rangle \right)^2.
\end{eqnarray}

Next we choose  ${\frak  g}\simeq {\bf C}_r $ series.
It is well known that the exterior products of $R(\omega _1) $ generically
are reducible, namely:
\begin{eqnarray}\label{eq:Al-12}
\wedge^k R(\omega _1) = R(\omega _k)\oplus \wedge^{k-2} R(\omega _1),
\nonumber\\
\omega _k = e_1 + e_2 + \dots + e_k, \qquad \mbox{for} \; k=2,\dots, r,
\end{eqnarray}
but the relations (\ref{eq:Al-2}), (\ref{eq:Al-3}) and (\ref{eq:Al-5})
hold also for ${\frak g}\simeq {\bf C}_r $ with $k=2,\dots,r $. Equation
(\ref{eq:Al-12}) reflects the existence of a nontrivial invariant subspace
in $\wedge^2 R(\omega _1) $, see \cite{Bour}:
\begin{equation}\label{eq:Cr-c}
|c\rangle=\sum_{k=1}^{N} (-1)^{k+1} |\gamma _k\wedge \gamma_{\bar{k}}
\rangle =\sum_{k,m=1}^{N} S_{km}|\gamma _k\wedge \gamma _m \rangle,
\end{equation}
where $S $ is the matrix entering into the definition of the symplectic
group (see (\ref{eq:S-mat}) below); namely
\begin{equation}\label{eq:Cr-def}
X\in Sp(2r) \qquad \leftrightarrow \qquad SX^T S^{-1}=X^{-1}.
\end{equation}
It is easy to check that due to (\ref{eq:Cr-def}) we have
$X|c\rangle=|c\rangle $ for any element $X\in Sp(2r) $.
Thus we establish that $c $ determines the one-dimensional invariant
subspace in $\wedge ^2R(\omega _1)=(\bbbc |c\rangle )\oplus R(\omega _2) $.

Let us now analyze the weights in the weight systems $\Gamma (\omega _k)
$ and their multiplicities. The highest weight $e_1+\dots +e_k $ has
multiplicity $ 1 $; the corresponding set of indices is
$I=\{1,2,\dots,k\} $. Let us consider next the weight $\gamma_{(1)}
=e_1+\dots+e_{k-2} $; to it there corresponds each of the following sets
of indices $I=\{1,2,\dots,k-2,p,\bar{p}\} $, $k-1\leq p\leq r $. Therefore
to $\gamma_{(1)}$ there corresponds the subspace $V(\gamma _{(1)})\subset
\wedge^kR(\omega _1) $ which is spanned by the vectors
$|\gamma_1\wedge\dots \wedge \gamma_{k-2}\wedge \gamma_p\wedge
\gamma_{\bar{p}} \rangle $ and has dimension $\dim V(\gamma_{(1)})=r-k+2$.
At the same time the Freudenthal formula shows that the multiplicity of
$\gamma _{(1)}=e_1+\dots+e_{k-2}  $ in $R(\omega _k) $ is $r-k+1$. This
difference is due to the fact that in  $V(\gamma _{(1)}) $ there exist a
one-dimensional invariant subspace determined by $|\gamma _1\wedge\dots
\wedge \gamma _{k-2}\wedge c \rangle $. The same argument can be applied
to each of the weights $\gamma_{(1)}'=w\gamma _{(1)} $ where $w $ is
an element of the Weyl group. Indeed, it is known that the Weyl group
preserves: i)~the lengths of the weights, i.e. $(\gamma_{(1)},
\gamma_{(1)}) =(\gamma_{(1)}',\gamma_{(1)}') =k-2 $;
ii)~the multiplicities of the weights.
In fact the Weyl group is isomorphic to the group of permutations
${\cal S}_{2r} $ of the indices $\{1,2,\dots,\bar{2},\bar{1} \}$;
therefore instead of looking at the transformed weight $\gamma
_{(1)}' $ we may consider the corresponding set of indices
which can be obtained from $I $ by applying a specific element of
${\cal S}_{2r} $. This analysis can be continued also by considering
weights of length $k-4 $, e.g. $\gamma _{(2)}=e_1+\dots e_{k-4} $.
Skipping the details we formulate the result, namely
\begin{equation}\label{eq:ast2}
\wedge ^{k} R(\omega_1) =R(\omega _k)\oplus (\bbbc |c\rangle )\wedge
(\wedge ^{k-2} R(\omega _1) ).
\end{equation}
It remains only to note that from (\ref{eq:Al-5}), (\ref{eq:ast2}) and
from  $V|c\rangle=|c\rangle $ there follows
\begin{eqnarray}\label{eq:ast3}
\langle \omega _k|V|\gamma _{i_1}\wedge \dots \wedge \gamma _{i_k}\rangle
&\equiv & V\left\{ \begin{array}{ccc} 1 & \dots & k \\ i_1 & \dots & i_k
\end{array}\right\} \nonumber\\
&=& \langle \omega _k |V|\gamma _I^{(k)} \rangle + \rho\langle
\omega _k |V|c\wedge\gamma _{I'} \rangle \nonumber\\
&=& \langle \omega _k |V|\gamma _I^{(k)} \rangle,
\end{eqnarray}
where $\rho $ is some constant, $\gamma _I^{(k)}\in R(\omega _k) $ and
$\gamma _{I'}\in \wedge ^{k-2} R(\omega _{1}) $. In the last line we
used also the fact that the representations $R(\omega _k) $ and
$\wedge^{k-2} R(\omega _{1}) $ span mutually orthogonal subspaces of
$\wedge^k R(\omega _1) $.

{}Finally let  ${\frak g }\simeq {\bf D}_r $ series. Then we have:
\numparts\label{eq:Al-13}
\begin{eqnarray}\label{eq:Al-13a}
\wedge^k R(\omega _1) &=& R(\omega _k), \qquad \mbox{for} \;
k=1,2,\dots, r-2 \\
\label{eq:Al-13b}
\wedge^{r-1} R(\omega _1) &=& R(\omega _{r-1}+\omega _r),\\
\label{eq:Al-13c}
\wedge^{r} R(\omega _1) &=& R(2\omega _{r})\oplus R(2\omega _{r-1}),
\end{eqnarray}
\endnumparts
where
\numparts\label{eq:Al-13om}
\begin{eqnarray}\label{eq:Al-13oma}
\omega _k= e_1+e_2+\dots + e_k, \qquad \mbox{for}\; k=1,2,\dots, r-2,\\
\label{eq:Al-13omb}
\omega _{r-1} = \case{1}{2} (e_1 + e_2 +\dots + e_{r-1} -e_r), \\
\label{eq:Al-13omc}
\omega _r = \case{1}{2} (e_1 + e_2 +\dots + e_{r-1} +e_r)
\end{eqnarray}
\endnumparts
{}From (\ref{eq:Al-13a}) and (\ref{eq:Al-13oma}) we find that the relations
(\ref{eq:Al-5}) hold also for ${\frak  g}\simeq {\bf D}_r $ with
$k=1,2,\dots,r-2 $. Let us now analyze the spinor representations
$R(\omega _{r-1}) $ and $R(\omega _{r}) $ of ${\bf D}_r $. It is known
\cite{VinOni} that:
\numparts\label{eq:Al-14}
\begin{eqnarray}\label{eq:Al-14a}
R(\omega _r)\otimes R(\omega _{r-1}) &=& R(\omega _{r-1} + \omega _r)
\oplus \sum_{i=1}^{[(r-1)/2]} R(\omega _{r-2i-1}), \\
\label{eq:Al-14b}
R(\omega _r)\mathop{\otimes}\limits_S R(\omega _{r}) &=& R(2 \omega _r)
\oplus \sum_{i=1}^{[r/4]} R(\omega _{r-4i}).
\end{eqnarray}
\endnumparts
where $\mathop{\otimes}\limits_S  $ means the symmetrized tensor product
and $R(\omega _0) $ is the trivial $1 $-dimensional representation of $
{\frak  g} $.

The basis in $R(\omega _{r}) \otimes R(\omega _{r-1}) $ is determined by
$|\gamma _I^{(r)} \otimes \gamma _{I'}^{(r-1)}\rangle  $ where $\gamma
_I^{(r)} \in \Gamma (\omega _r)$ and $\gamma _{I'}^{(r-1)}\in \Gamma
(\omega _{r-1}) $.  The sets of indices $I=\{i_1,\dots,i_r\} $ and
$I'=\{i_1',\dots,i_r'\} $ are related to the weights $\gamma _I^{(r)} $
and $\gamma _{I'}^{(r-1)} $ by
\begin{equation}\label{eq:Al-16}
\fl \gamma_{I'}^{(r-1)} = \case{1}{2} \sum_{k=1}^{r} \sigma' _k
e_k = \case{1}{2} \sum_{k=1}^{r} \gamma _{i_k'}, \qquad \gamma _{I}^{(r)}
= \case{1}{2} \sum_{k=1}^{r} \sigma _k e_k =\case{1}{2} \sum_{k=1}^{r}
\gamma _{i_k},
\end{equation}
where $\sigma _k $ and $\sigma _k' $ take values $\pm 1 $ and
\begin{equation}\label{eq:Al-16'}
\prod_{k=1}^{r}\sigma _k' =-1 \qquad  \mbox{and}
\qquad \prod_{k=1}^{r}\sigma _k =1.
\end{equation}
Let us now consider a pair of weights $\gamma _I^{(r)} $ and
$\gamma_{I'}^{(r-1)}$ such that $\sigma _k=\sigma _k' $ for all but one
value of $k $, i.e.  $I\cap I' =\{j_1,\dots,j_{r-1}\} $. Then the vectors
$\gamma_{I}^{(r)}+\gamma _{I'}^{(r-1)}$ and $\omega _r +\omega _{r-1} $
have the same length and one can check that they are related by a Weyl
group transformation. Therefore $|\gamma _{I}^{(r)}\otimes \gamma _{I
'}^{(r-1)}\rangle  $ belongs to the module $R(\omega _{r-1}+\omega _r)  $
in (\ref{eq:Al-14a}) and
\numparts\label{eq:Al-15}
\begin{eqnarray}\label{eq:Al-15a}
 V\left\{ \begin{array}{cccc} 1 & 2 & \dots & r-1\\ i_1 & i_2 &
\dots & i_{r-1} \end{array} \right\} &=&
\langle \omega _{r-1} +
\omega _r|V| \gamma _{I'}^{(r-1)} + \gamma _{I}^{(r)}  \rangle \nonumber\\
&=& \langle \omega _{r-1} |V| \gamma _{I'}^{(r-1)} \rangle \langle
\omega _{r} |V| \gamma _{I}^{(r)}\rangle.
\end{eqnarray}
{}Finally, the spinor representation $R(\omega _r) $ is considered quite
analogously to the one of the ${\bf B}_r $ case. There is only one
additional fact that should be taken care of: now the weights $2\omega _r $
and $2\omega _{r-1} $ have the same length but are related by an outer
automorphism rather than by a Weyl group element.  In order to separate
the weights belonging to the modules $R(2\omega _{r}) $ and $R(2\omega
_{r-1}) $ we need projectors onto each of these invariant subspaces in
$\wedge^r R(\omega _r) $ in (\ref{eq:Al-13c}). As we shall see below we
will need to sort out only the weights of length $r $ which are given by
either $2\gamma _I^{(r-1)} $ or by $2\gamma _I^{(r)} $. The projectors
which will separate these two types of weights can be constructed by
using (\ref{eq:Al-16'}) and the fact that $\zeta _{\bar{k}} = - \zeta_k$;
the matrix elements of these projectors are given by:
\begin{eqnarray}\label{eq:Al-15c}
f_{I}^\pm \equiv f^\pm_{i_1,i_2,\dots,i_r} = \case{1}{2} \left( 1 \pm {
 \zeta _1 \zeta _2\dots \zeta _r \over \zeta _{i_1} \zeta _{i_2} \dots
\zeta _{i_r}} \right).
\end{eqnarray}
Indeed, it is easy to check now that
\begin{equation}\label{eq:Al-16''}
\fl f_I^+ \gamma _{I'}^{(r-1)} =f_{I'}^- \gamma _{I}^{(r)} =0, \qquad
f_{I}^+ \gamma _{I}^{(r)} = \gamma _{I}^{(r)}, \qquad
f_{I'}^- \gamma_{I'}^{(r-1)} = \gamma_{I'}^{(r-1)}.
\end{equation}
Thus we obtain the relation:
\begin{eqnarray}\label{eq:Al-15b}
\fl f^+_{i_1,i_2,\dots,i_r} V\left\{
\begin{array}{cccc} 1 & 2 & \dots & r\\ i_1 & i_2 & \dots & i_{r}
\end{array} \right\}= \langle 2 \omega _r|V| 2\gamma _{I}^{(r)}\rangle =
(\langle \omega _{r} |V| \gamma_{I}^{(r)} \rangle)^2.
\end{eqnarray}
\endnumparts
{}From (\ref{eq:Al-15a})--(\ref{eq:Al-15b}) we find the necessary
expressions for $\langle \omega _{r-1} |V| \gamma_{I}^{(r-1)} \rangle $
and $\langle \omega _{r} |V| \gamma_{I}^{(r)} \rangle $ through the minors
of $V $, see (\ref{eq:ap-4.1}).

\section{The solutions of the CTC revisited}
We will now analyze the structure of the solutions to the CTC for each of
the classical series of Lie algebras. Our aim will be to write them
down in the form (\ref{eq:**a}), (\ref{eq:**b}) and calculate the functions
$W^{(k)}(\vec{\zeta },\gamma ) $ for each of the classical series
separately.

\subsection{The ${\bf A}_r $ series.}

It is well known \cite{Moser,OlshPer,Toda,And} that the solutions in this
case can be expressed through the principle minors of the group element
$G(t) $ (\ref{eq:1-1}). Using the Binet-Cauchy formula, or equivalently
(\ref{eq:Al-5}) and (\ref{eq:ap-2.1}) we obtain the functions:
\begin{eqnarray}\label{eq:A1}
A_1(t)= \sum_{k=1}^{N} r_k^2 \rme ^{-2\zeta _kt},\\
\label{eq:Ak}
\fl A_k(t)= \sum_{i_1<i_2<\dots <i_k}^{} r_{i_1}^2 r_{i_2}^2 \dots
r_{i_k}^2\rme ^{-2(\zeta _{i_1} + \zeta _{i_2} \dots + \zeta_{i_k} )t}
W^2(i_1,i_2,\dots, i_k),\\
\label{eq:AN}
A_N(t)= \prod_{s=1}^{N} r_s^2 W^2(1,2,\dots,N)= \rme ^{-Nq_1(0)}.
\end{eqnarray}
which are proportional to the above-mentioned principle minors of $G(t) $.
Here $\zeta _k $, $r_k $ are the scattering data of the Lax matrix $L(0) $
introduced in (\ref{eq:Tau})--(\ref{eq:norm1}) and $W_I(\vec{\zeta})
\equiv W(i_1,i_2,\dots,i_k) $ is the Vandermonde
determinant:
\begin{equation}\label{eq:WI}
W_I(\vec{\zeta })\equiv W(i_1,i_2,\dots,i_k) = \prod_{s>p;\,
s,p\in I}^{} 2(\zeta _{s} -\zeta _{p}).
\end{equation}
Note that each of the  factors in the right hand side of (\ref{eq:WI}) can
be viewed as the scalar product $(\vec{\zeta },\alpha ) $ where $\alpha  $
is an appropriately chosen root $\alpha \in \Delta _{{\frak  g}} $. This
and the fact that $(\vec{\zeta },w_0(\alpha) )  =(w_0(\vec{\zeta }),\alpha
) $ make it easy to introduce in a natural way the action of the Weyl
group element $w_0 $ on $W_I(\vec{\zeta }) $, namely $w_0:  W_I(\vec{\zeta
})\to W_I(w_0(\vec{\zeta })) $.

The solution of the corresponding CTC is then given by:
\begin{equation}\label{eq:q-om}
(\vec{q}(t)-\vec{q}(0), \omega _k) = \ln A_k(t)
\end{equation}
or
\begin{equation}\label{eq:qk}
q_k(t) = q_1(0) + \ln {A_k(t)  \over A_{k-1}(t) }
\end{equation}
with $A_0=1 $ and $q_1(0) $ defined by (\ref{eq:AN}).

{}For further convenience we introduce the functions
\begin{equation}\label{eq:B-k}
B_k(t) = \rme ^{k q_1(0)} A_k(t),
\end{equation}
and rewrite the solution to the CTC in the form:
\begin{equation}\label{eq:q-omB}
(\vec{q}(t), \omega _k) = \ln B_k(t)
\end{equation}

Introducing now the set of variables (\ref{eq:tau-Ar}) and taking into
account (\ref{eq:Al-1a}) we easily cast the solution
(\ref{eq:q-omB}) into the form (\ref{eq:**a}), (\ref{eq:**b}) with
\begin{eqnarray}\label{eq:24.2}
\fl B_k(t)={\cal B}_k(t)&=&  \sum_{\gamma _{I}^{(k)} \in \Gamma (\omega
_k)} \exp\left(\vec{\tilde{\varphi} }(t),\gamma _{I}^{(k)} \right)
W^2_I(\vec{\zeta }) (W(1,2,\dots,N))^{-2k/N},\\
\label{eq:24.3}
&& \vec{\tilde{\varphi} }(t) = \left(\tilde{\varphi} _1(t),\dots,
\tilde{\varphi }_N(t)\right), \qquad  \tilde{\varphi }_k(t)=-2\zeta _kt +
\tilde{\varphi }_{0k},
\end{eqnarray}
where $\tilde{\varphi}_{0k} $ were introduced in (\ref{eq:tau-Ar}). Another
possibility is to use $\vec{\varphi}(t)=-2\vec{\zeta }t
+\vec{\varphi}_{0}$ with
\begin{eqnarray}\label{eq:phi-0k}
\varphi _{0k} = \tilde{\varphi }_{0k} - \ln w_k - {2  \over  N} \ln
W(1,\dots,N) = q_1(0) + \ln {r_k^2  \over w_k }, \nonumber\\
w_k = \prod_{s=1}^{k-1} {1  \over 2(\zeta _k -\zeta
_s) } \prod_{s=k+1}^{N} {1  \over 2(\zeta _s - \zeta _k)  }.
\end{eqnarray}
which gives:
\begin{equation}\label{eq:B-A-k}
\fl {\cal  B}_{{\bf A}_r;k}(t) = \sum_{i_1<i_2<\dots <i_r}^{N}
\rho _{i_1}^2 \rho _{i_2}^2\dots \rho _{i_k}^2 \rme^{-2(\zeta_{i_1}+
\dots +\zeta_{i_k})t } W_I^2(\vec{\zeta }) w_{i_1}\dots w_{i_k},
\end{equation}
with $\rho _k =\rme^{\varphi _{0k}} $. Then finally ${\cal B}_{{\bf
A}_r;k} (t) $ can be rewritten as:
\begin{equation}\label{eq:Bk-A}
{\cal  B}_{{\bf A}_r;k}(t)= \sum_{\gamma _{I}^{(k)} \in \Gamma (\omega
_k)} \exp\left( \vec{\varphi }(t),\gamma _{I}^{(k)} \right)
W^{(k)}(\vec{\zeta },\gamma _I^{(k)}),
\end{equation}
with
\begin{equation}\label{eq:Ar-Wk}
\fl W^{(1)}(\zeta,\gamma _k) = w_k, \qquad
W^{(k)}(\vec{\zeta },\gamma _I^{(k)}) = W_I^2(\vec{\zeta}) \prod_{s\in
I}^{}w_s, \quad k=2,3,\dots,r.
\end{equation}
Now it is more natural to consider
\begin{equation}\label{eq:tau-Ar1}
{\cal  T}_{{\bf A}_r} \equiv \{ \zeta _1,\dots, \zeta _N, \varphi _{01},
\dots, \varphi _{0N} \}, \qquad N=r+1
\end{equation}
rather than (\ref{eq:tau-Ar}), as the minimal set of scattering data for
the ${\bf A}_r $ algebra case.  Indeed, the elements of
(\ref{eq:tau-Ar1}), like the ones of (\ref{eq:tau-Ar}), naturally satisfy
the identities:
\begin{equation}\label{eq:Ar-id}
\sum_{k=1}^{N} \zeta _k = (\vec{\zeta },\vec{\epsilon }) =0, \qquad
\sum_{k=1}^{N} \varphi _{0k}= (\vec{\varphi },\vec{\epsilon }) =0,
\end{equation}
$\epsilon =e_1+\dots +e_N $, which is directly related to the fact, that
all the roots of ${\bf A}_r $ also satisfy $(\alpha,\vec{\epsilon }) =0
$.

However (\ref{eq:tau-Ar1}) has also the advantage that imposing on its
elements the symmetry condition (\ref{eq:26a}) and
\begin{equation}\label{eq:226a}
\varphi _{0k} = - \varphi _{0\bar{k}}, \qquad \bar{k}=N+1-k,
\end{equation}
one can get the minimal set of scattering data for the ${\bf B}_r $ and
${\bf C}_r $ series.

Indeed, let us choose ${\frak  g}\simeq {\bf A}_{2r} $ and let us impose
on ${\cal  T}_{{\bf A}_{2r}} $ (\ref{eq:26a}), (\ref{eq:226a}). First
note, that from (\ref{eq:26a}) and (\ref{eq:phi-0k}) with $N=2r+1 $ we
immediately find that $w_k $ takes the form (\ref{eq:w-k-b}) for
$k=1,\dots, r $ and (\ref{eq:r0-b}) for $k=r+1 $; besides one can check
that $w_{\bar{k}} = w_k $.

Next the condition (\ref{eq:226a}) with $\varphi _{0,\bar{k}} =q_1(0) + \ln
(r_{\bar{k}}^2/w_k)$ leads immediately to the relation (\ref{eq:26b}).
Expressing from it $q_1(0) $ as $\ln(w_k/(r_kr_{\bar{k}})) $ we easily get
$\varphi _{0,k}=\ln(r_k/r_{\bar{k}}) $ for $k=1,\dots,r $ and $\varphi
_{0,r+1}=0 $. Thus we showed that ${\cal  T}_{{\bf A}_{2r}} $ with
(\ref{eq:26a}) and (\ref{eq:226a}) reduces to (\ref{eq:Tau-g}) for the
series ${\bf B}_{r}$.

The same procedure applied to ${\cal  T}_{{\bf A}_{2r-1}} $  provides
(\ref{eq:Tau-g}) for the series ${\bf C}_{r}$.

\subsection{The ${\bf B}_r $ series.}

The solution is obtained from the one for ${\bf A}_{2r} $ series by
imposing the relations (\ref{eq:26a})--(\ref{eq:r0-b}). Due to this only
the first $r $ of the functions $A_k(t) $ in (\ref{eq:q-om}) are
independent. From the analysis in Sect.~2 we see that only the expression
for $A_r(t) $ requires additional special treatment\footnote{This is
related to the existence of the spinor representation.}, see
(\ref{eq:Al-10}). Thus we find:
\begin{eqnarray}\label{eq:49.4a}
A_k(t)&=& \rme ^{-kq_1(0)} {\cal  B}_k(t), \qquad  \mbox{for
$k=1,\dots,r-1 $} \\
\label{eq:49.4b}
A_r(t)&=& \rme ^{-rq_1(0)} {\cal  B}_r^2(t).
\end{eqnarray}
where ${\cal  B}_k(t) $ are of the form (\ref{eq:**b}),
\begin{equation}\label{eq:50.2a}
{\cal  B}_k(t) = \sum_{\gamma _I^{(k)} \in \Gamma (\omega _k)}^{} \exp
\left( \vec{\varphi} (t), \gamma _I ^{(k)}\right) W^{(k)}(\vec{\zeta },
\gamma _I^{(k)}),
\end{equation}
with
\begin{equation}\label{eq:50.4a}
\fl W^{(1)}(\zeta,\gamma _k) = w_k, \qquad
W^{(k)}(\vec{\zeta },\gamma _I^{(k)})= W_I^2 (\vec{\zeta }) \prod_{s\in
I}^{} w_s, \qquad \vec{\varphi }(t)=-2 \vec{\zeta }t + \vec{\varphi }_0,
\end{equation}
for $k=1,\dots,r-1 $ and
\begin{equation}\label{eq:50.4b}
W^{(r)}(\vec{\zeta }, \gamma _I^{(r)})= W_I (\vec{\zeta })
\prod_{s=1}^{r} w_s^{1/2}.
\end{equation}
Here $W_I(\vec{\zeta }) $ is the Vandermonde determinant (\ref{eq:WI}).
We remind that to each weight $\gamma _I^{(k)}\in \Gamma (\omega _k) $
there corresponds an ordered set of indices $I=\{i_1 <\dots< i_k \}$ and
that $w_k $ are defined by (\ref{eq:w-k-b}), (\ref{eq:r0-b}). Each set $I
$ uniquely defines $\gamma _I^{(k)} $. If the weight $\gamma _I^{(k)} $ has
multiplicity greater than $1 $ then there exist several sets of indices
related to $ \gamma _I^{(k)} $. For example, the weights $e_1+\dots + e_k
$ and $e_1 +\dots +e_{k-1} $ in $\Gamma (\omega _k) $ have multiplicities
$1 $; the corresponding sets of indices are $\{1,2,\dots,k\} $ and
$\{1,2,\dots,k-1,r+1\} $ respectively. The weight $e_1+\dots+e_{k-2} $
however has multiplicity $r-k+2 $; one may assign to it each of the
following sets of indices $ \{1,2,\dots,k-2,p,\bar{p}\}$ where $k-1 \leq p
\leq r$. Analogously to the weight $e_1+\dots+e_{k-3} $ and $
e_1+\dots+e_{k-4}  $ with multiplicities $r-k+3 $ and
$\left(\begin{array}{c} r-k+4\\ 4 \end{array} \right)$ one can assign each
of the sets $\{ 1,\dots,k-3,p,r+1,\bar{p}\} $ with $k-2\leq p\leq r $ and
$\{ 1,\dots,k-4, p_1,p_2,\bar{p}_1,\bar{p}_2\} $ with $k-3 \leq p_1<p_2\leq
r$. In other words we find that the number of sets $I $ corresponds to the
number of weights provided each weight is counted as many times as
its multiplicity.

We note that ${\cal  B}_k(t) $ can also be written in the form:
\begin{equation}\label{eq:50.2b}
\fl {\cal  B}_k(t)= \sum_{\gamma _I^{(k)}>0}^{} 2\cosh \left(\vec{\varphi
}(t),\gamma_I^{(k)}\right) W^{(k)}(\vec{\zeta },\gamma _I^{(k)}) +
\sum_{I:\gamma _I^{(k)}=0} W^{(k)}(\vec{\zeta },\gamma _I^{(k)}),
\end{equation}
where the second sum runs over the sets of indices corresponding to the
weight equal to zero. For the ${\bf B}_r $ series these sets are of the
form:
\[ \{
p_1,\dots,p_s,\bar{p}_1,\dots,\bar{p}_s\} \qquad \mbox{for $k=2s $},\] and
\[ \{ p_1,\dots,p_s, r+1,\bar{p}_1,\dots,\bar{p}_s\} \qquad \mbox{for
$k=2s+1 $}.\]

{}Formula (\ref{eq:50.2b}) reflects the fact that $\Gamma (\omega _k) $ is
symmetric in the sense that if $\gamma_I^{(k)} \in\Gamma (\omega _k) $ then
$-\gamma_I^{(k)}=w_0(\gamma_I^{(k)} ) \in\Gamma (\omega _k) $. One can
also check that
\[
\fl W^{(k)}(\vec{\zeta },\gamma _I^{(k)}) =  W^{(k)}(w_0(\vec{\zeta }),
w_0(\gamma _I^{(k)})) = W^{(k)}(\vec{\zeta },\gamma _{\bar{I}}^{(k)}),
\qquad \bar{I}=\{ \bar{i}_k, \dots, \bar{i}_1\}.
\]

\subsection{The ${\bf C}_r $ series.}
\begin{remark}\label{rem:Cr}
When one imposes the symmetry condition (\ref{eq:26a}), (\ref{eq:26b})
on the ${\bf A}_{2r-1} $ CTC the corresponding system of equations is
slightly different from (\ref{eq:gCTC}). The difference consists only in
the coefficient of the term $\rme^{-(\vec{q},\alpha _r)} $ which comes out
as $\alpha _r/2 $ instead of $\alpha _r $. The extra $1/2 $ factor is easy
to take into account and therefore for the ${\bf C}_r $-series the
relation between $q_k(t) $ and ${\cal  B}_k(t) $ is slightly different,
namely:
\begin{equation}\label{eq:Cr-re}
q_k(t) = \ln {{\cal  B}_k(t)  \over {\cal  B}_{k-1}(t) } + {1 \over 2}\ln
2.
\end{equation}
\end{remark}

Now we insert (\ref{eq:w-k-c}) into (\ref{eq:A1})--(\ref{eq:Ak}) to get
the solutions for the ${\bf C}_r $ series. Thus we get
\begin{equation}\label{eq:Cr-}
A_k(t)=\rme^{-kq_1(0)} B_k(t),
\end{equation}
for all $k=1,2,\dots,r $ where
\begin{eqnarray}\label{eq:57.3}
\fl B_k(t) = \sum_{i_1<\dots <i_k}^{} \exp \left( \vec{\varphi}(t),
\gamma _{i_1\dots i_k}\right) W^{(k)}(\vec{\zeta },
\gamma _{i_1\dots i_k}), \qquad \vec{\varphi }(t)=-2\vec{\zeta }t
+\vec{\varphi }_0,\\
\label{eq:57.4}
\fl W^{(1)}(\zeta,\gamma _k) = w_k, \qquad W^{(k)}(\vec{\zeta },
\gamma_{i_1\dots i_k}) = W^2(i_1,\dots, i_k)\prod_{s=1 }^{k} w_{i_s}.
\end{eqnarray}
and $\gamma _{i_1\dots i_k} $ is a weight in $\wedge^k R(\omega _1) $.
{}From our analysis in Section~2 (see (\ref{eq:ast2}) each of the weights
$\gamma_{i_1,\dots, i_k}=\gamma _{i_1}+\dots + \gamma_{i_k}$ can be split
into
$|\gamma_{i_1,\dots,i_k}\rangle = |\gamma _I^{(k)}\rangle +\rho |c\wedge
\gamma_{I'}\rangle $ \
(see (\ref{eq:ast3})) and we have
\begin{eqnarray}\label{eq:Cr-*}
\langle \omega _k |V|\gamma_{i_1,\dots,i_k }\rangle &\equiv &
V\left\{ \begin{array}{ccc} 1 & \dots & k \\ i_1 & \dots & i_k
\end{array}\right\} = \langle \omega _k|V|\gamma^{(k)} _{I}\rangle.
\end{eqnarray}
This means that the summation over all sets of indices $i_1<i_2<\dots
<i_k $ in (\ref{eq:57.3}) reduces to the sum over the weights of $\Gamma
(\omega _k) $ only:
\begin{equation}\label{eq:57.3a}
{\cal  B}_k(t)= B_k(t)=\sum_{\gamma _I^{(k)}\in \Gamma (\omega _k)}^{} \exp
\left(\vec{\varphi }(t), \gamma _I^{(k)}\right) W^{(k)}(\vec{\zeta },
\gamma _{I}^{(k)}).
\end{equation}
\begin{equation}\label{eq:slgvslkhgbel}
 W^{(k)}(\vec{\zeta },\gamma _I^{(k)})= W_I^2 (\vec{\zeta }) \prod_{s\in
I}^{} w_s, \qquad k=2,3,\dots, r
\end{equation}
i.e. we cast the solution in the form (\ref{eq:**a}), (\ref{eq:**b}).
Besides, like for the ${\bf B}_r $-series again $w_0(\gamma _I^{(k)})= -
\gamma _I^{(k)} $ so we have:
\begin{equation}\label{eq:57.5}
\fl {\cal  B}_k(t)=\sum_{\gamma _I^{(k)}>0} 2\cosh \left(\vec{\varphi}(t),
\gamma_I^{(k)}\right) W^{(k)}(\vec{\zeta },\gamma _{I}^{(k)}) +
\sum_{I:\gamma _{I}^{(k)}=0} W^{(k)}(\vec{\zeta },\gamma _{I}^{(k)}).
\end{equation}
We note that the sets of indices corresponding to the $\gamma
_{I}^{(k)}=0$ are given by $\{ p_1,\dots,p_s,\bar{p}_1,\dots,\bar{p}_s\} $
for $k=2s $ and by the empty set for $k=2s+1 $.

\subsection{The ${\bf D}_r $ series.}

There are some differences in treating this case due to the fact that the
Lax matrix $L(0) $ (\ref{eq:ap-3.2}) is not a tri-diagonal one.
Nevertheless Moser formula (\ref{eq:A1})--(\ref{eq:AN}) for $N=2r $,
together with the corresponding involution (\ref{eq:26a}), (\ref{eq:26b}),
(\ref{eq:w-k-d}) provides the solution to the ${\bf D}_r $. Due to the
somewhat different structure of the eigenmatrix $V $ we find that the
first $r-1 $ functions $A_k(t) $ are given by (\ref{eq:A1}), (\ref{eq:Ak})
and only $A_r(t) $ must be replaced by \cite{GEI}:
\begin{equation}\label{eq:Dr-Ar}
\fl\tilde{A}_r(t) = \sum_{i_1<\dots <i_r}^{} r_{i_1}^2 \dots r_{i_r}^2
\exp \left(-2 (\zeta _{i_1}+\dots +\zeta _{i_r})t \right)
(f^+_{i_1,\dots,i_r})^2 W^2(i_1,\dots,i_r).
\end{equation}

Besides, the ${\bf D}_r $ algebras have two spinor representations which
requires additional care.  Note also that due to (\ref{eq:ap-4.1}) the
projector $f_{i_1\dots i_r}^+ $ (\ref{eq:Al-15c}) enters in a natural
way into $\tilde{A}_r $. Thus in the right hand side only the terms
related to the weights of $R(2\omega _{r}) $ give non-vanishing
contributions.

Let us introduce the variables (\ref{eq:Tau-g}) and along with
(\ref{eq:B-k}) for $k=1,\dots, r-1 $ let us put $B_r(t)=\rme ^{r q_1(0)}
\tilde{A}_r(t) $; then we can rewrite the solution for the ${\bf D}_r
$-series in the form:
\begin{eqnarray}\label{eq:Dr-2a}
(\vec{q}(t),\omega_k) &=& \ln B_k(t), \qquad \mbox{for}\; k=1,\dots,r-2,\\
\label{eq:Dr-2b}
(\vec{q}(t),\omega _{r-1} + \omega _r) &=& \ln B_{r-1}(t), \qquad
(\vec{q}(t),2\omega _{r} ) = \ln B_{r}(t).
\end{eqnarray}
Our analysis in Section~2 shows that
$w_k=w_{\bar{k}} $ and
\begin{eqnarray}\label{eq:Dr-Br}
B_k(t)&=& {\cal  B}_k(t), \qquad \mbox{for}\; k=1,\dots, r-2, \nonumber\\
B_{r-1}(t) &=& {\cal  B}_{r-1}(t) {\cal  B}_{r}(t), \qquad
B_{r}(t) = {\cal  B}_{r}(t)^2,
\end{eqnarray}
with
\begin{eqnarray}\label{eq:Dr-Br-}
\fl {\cal  B}_{k}(t) &=& \sum_{\gamma_{I}^{(k)} \in \Gamma (\omega
_{k})}^{} \exp \left( \vec{\varphi }(t),\gamma_{I}^{(k)} \right) W^{(k)}
(\vec{\zeta }, \gamma _{I}^{(k)}), \qquad \mbox{for}\; k=1,\dots, r,
\end{eqnarray}
Here:
\begin{equation}\label{eq:asdgasdg}
W^{(1)}(\zeta,\gamma _k) = w_k,
\qquad \vec{\varphi }(t)=-2 \vec{\zeta }t + \vec{\varphi }_0,
\end{equation}
\begin{eqnarray}\label{eq:Dr-Wk}
&&W^{(k)}(\vec{\zeta },\gamma _I^{(k)}) = W_I^2(\vec{\zeta})
\prod_{s\in I} w_s, \qquad  k=1,\dots,r-2,\\
\label{eq:Dr-Wr-a}
&& W^{(r-a)}(\vec{\zeta },\gamma _I^{(r-a)}) = \sqrt{w_1\dots w_r}
W_I(\vec{\zeta}), \qquad a=0,1,
\end{eqnarray}
and the set of indices $I =\{i_1,\dots,i_{r}\}$ in (\ref{eq:Dr-Wr-a}) is
such that it determines uniquely the weight $\gamma_I^{(r-a)} \in \Gamma
(\omega _{r-a}) $ in the corresponding spinor representation. In addition
\begin{equation}\label{eq:Dr-omr-a}
(\vec{q}(t),\omega_{r-a}) = \ln {\cal  B}_{r-a}(t).
\end{equation}

The explicit solutions for RTC with the simplest choices for ${\frak  g} $
with rank 2 in invariant form were proposed in the monograph \cite{LeSav};
they coincide with the particular cases of the ones given above provided
a proper identification of the variables is performed. Here we choose to
provide as examples the solution to the ${\bf D}_4 $ case and its
relation to the solutions for ${\bf B}_3 $ and ${\bf G}_2 $.

\begin{example}\label{exa:1}
Let ${\frak  g}\simeq {\bf D}_4 =so(8) $. Then it has three 8-dimensional
representations: $R(\omega _1) $ and the two spinor ones $R(\omega _3) $
and $R(\omega _4) $. The representation $R(\omega _2) $ is of dimension
28. We remind also that ${\bf D}_4 $ has an outer automorphism $v_1 $ of
order 3 which interchanges the $8 $-dimensional representations; more
precisely:
\begin{eqnarray}\label{eq:D4-V}
v_1:\alpha _1 \to \alpha _3 \to \alpha _4 \to \alpha _1; \qquad
v_1\alpha _2=\alpha _2, \nonumber\\
v_1: \omega _1\to \omega _3 \to \omega _4 \to \omega _1; \qquad
v_1\omega _2=\omega _2.
\end{eqnarray}
The equations of motion for the ${\bf D}_4 $-CTC have the form
\begin{eqnarray}\label{eq:151.3}
q_{1,tt} = \rme^{q_2 -q_1}, \qquad  q_{2,tt} = \rme^{q_3 -q_2}- \rme^{q_2
-q_1}, \nonumber\\
\fl q_{3,tt} = -\rme^{q_3 -q_2} + \rme^{q_4 -q_3} +\rme^{-q_3 -q_4}, \qquad
q_{4,tt} = -\rme^{q_4 -q_3} + \rme^{-q_3 -q_4}.
\end{eqnarray}
The solution is provided by
\begin{eqnarray}\label{eq:151.3'}
q_1(t) &=& \ln {\cal  B}_1(t), \qquad
q_2(t) = \ln {{\cal  B}_2(t) \over {\cal  B}_1(t)}, \nonumber\\
q_3(t) &=& \ln {{\cal  B}_3(t){\cal  B}_4(t)\over {\cal  B}_2(t) }, \qquad
q_4(t) = \ln {{\cal  B}_4(t) \over {\cal  B}_3(t) },
\end{eqnarray}
where
\numparts\label{eq:149}
\begin{eqnarray}\label{eq:149.1}
{\cal  B}_1(t) = 2 \sum_{k=1}^{4} w_k \cosh \varphi _k(t), \\
\label{eq:149.2}
\fl {\cal  B}_2(t) = 8 \sum_{i<j}^{4} w_i w_j \left[ (\zeta _j-\zeta _i)^2
\cosh (\varphi _i +\varphi _j) + (\zeta _j+\zeta _i)^2 \cosh (\varphi _i
-\varphi _j) \right] \nonumber\\  + 16\sum_{i=1}^{4} w_i^2 \zeta _i^2,\\
\label{eq:149.3}
{\cal  B}_3(t) = 2^{-11} \prod_{i<j}^{4} {1  \over \zeta _i^2 -\zeta
_j^2} \left\{ W(1,2,3,\bar{4}) \cosh {\varphi _1+\varphi _2 +\varphi _3
-\varphi _4 \over 2} \right. \nonumber\\
\fl +  W(1,2,4,\bar{3}) \cosh {\varphi _1+\varphi _2 -\varphi _3
+\varphi _4 \over 2} + W(1,3,4,\bar{2}) \cosh {\varphi _1-\varphi _2
+\varphi _3 +\varphi _4 \over 2} \nonumber\\
\left. + W(2,3,4,\bar{1}) \cosh {\varphi_1-\varphi _2 -\varphi _3
-\varphi_4 \over 2} \right\},\\
\label{eq:149.4}
{\cal  B}_4(t) = 2^{-11} \prod_{i<j}^{4} {1  \over \zeta _i^2 -\zeta
_j^2} \left\{ W(1,2,3,4) \cosh {\varphi _1+\varphi _2 +\varphi _3
+\varphi _4 \over 2} \right. \nonumber\\
\fl +  W(1,2,\bar{4},\bar{3}) \cosh {\varphi _1+\varphi _2
-\varphi _3 -\varphi _4 \over 2} + W(1,3,\bar{4},\bar{2}) \cosh {\varphi
_1-\varphi _2 +\varphi _3 -\varphi _4 \over 2} \nonumber\\
\left. + W(2,3,\bar{4},\bar{1})\cosh {\varphi _1-\varphi _2 -\varphi _3
+\varphi _4 \over 2} \right\}, \\
\label{eq:149.4'} \varphi _k(t)  =-2\zeta _kt + \varphi _{0k}.
\end{eqnarray}
\endnumparts
We remind also that $\bar{4}=5 $, $\bar{3}=6 $, $\bar{2}=7$, $\bar{1}=8 $;
$W(i,j,k,l) $ are defined by (\ref{eq:WI}) and in both the summation
and the products, denoted above by $i<j $ we mean that $i $ and $j $ take
values from 1 to 4.
\end{example}

It is well known that the automorphism $v_1 $ maps not only the
fundamental weights $\omega _k $ as in (\ref{eq:D4-V}) but also the whole
sets of weights, e.g. $v_1: \Gamma (\omega _1)\to \Gamma (\omega _3)\to
\Gamma (\omega _4) $. Using the definition of ${\cal  B}_k(t) $ and the
properties of the scalar products $(\vec\zeta,v_1\gamma ^{(k)}) =
(v_1^{-1}\vec\zeta,\gamma ^{(k)})$ we obtain
\begin{eqnarray}\label{eq:149.9}
{\cal  B}_1(t;\vec{\zeta }, \vec{\varphi _0}) =
{\cal  B}_3(t;v_1^{-1}\vec{\zeta }, v_1^{-1}\vec{\varphi _0}) =
{\cal  B}_4(t;v_1\vec{\zeta }, v_1\vec{\varphi _0}), \nonumber\\
{\cal  B}_2(t;\vec{\zeta }, \vec{\varphi _0}) =
{\cal  B}_2(t;v_1\vec{\zeta }, v_1\vec{\varphi _0}).
\end{eqnarray}
These relations are compatible with (\ref{eq:**a}), (\ref{eq:**b}), i.e.,
\[
(\vec{q}(t),\omega_{3,4})=\ln {\cal B}_{3,4}(t;\zeta,\varphi_0)=
(\vec{q}(t),v_1^{\pm 1}\omega_{1})=\ln {\cal  B}_1(t;v_1^{\pm 1}\vec{\zeta
}, v_1^{\pm 1}\vec{\varphi _0}).
\]

\begin{example}\label{exa:B3}
Next we will derive the result for the ${\bf B}_3 $  starting from the
${\bf D}_4 $ case. It is well known that ${\bf B}_3 $ can be obtained from
${\bf D}_4 $ by imposing a symmetry condition with respect to the outer
automorphism $v_2 $ of ${\bf D}_4 $ defined by:
\begin{equation}\label{eq:154.1}
v_2\alpha _k =\alpha _k, \quad \mbox{for\;} k=1,2; \qquad
v_2\alpha _3=\alpha _4, \qquad v_2\alpha _4=\alpha _3.
\end{equation}
The symmetry with respect to $v_2 $ reflects on the scattering data of the
${\bf D}_4 $-case by
\begin{equation}\label{eq:154.2}
\zeta _4=0, \qquad \varphi _{04}=0, \qquad \mbox{or} \qquad \varphi
_4(t)=0.
\end{equation}
To the end of this example, when referring to the variables related to
${\bf D}_4 $ we will assume these  conditions imposed and will denote this
fact by additional ``prime''; the corresponding variables for the ${\bf
B}_3 $-case we will denote by the same letters with additional ``tilde''.
Then inserting (\ref{eq:154.2}) into (\ref{eq:w-k-d}) we can write:
\begin{equation}\label{eq:154.4}
w_k' = 2\tilde{w}_k, \qquad k=1,2,3; \qquad w_4'= \tilde{w}_4.
\end{equation}
In analogy with (\ref{eq:149.9}) we have
\begin{equation}\label{eq:1kjhnfclwk}
{\cal  B}'_4(t;\vec{\zeta }, \vec{\varphi _0}) =
{\cal  B}'_3(t;v_2\vec{\zeta }, v_2\vec{\varphi _0})
\end{equation}
and after imposing the $v_2$-involution symmetry, namely,
$(\vec{q}(t),\omega_{k})= (\vec{q}(t),v_2\omega_{k})$ for each $k$ we have
$q_4'=0$. Indeed using (\ref{eq:154.2}) one can find that
${\cal  B}'_4(t;\vec{\zeta }, \vec{\varphi _0}) =
{\cal  B}'_3(t;\vec{\zeta },\vec{\varphi _0}) $
which leads to $q_4'=0$. This way $q_1',q_2'$ and $q_3'$ give us solutions
of a system, equivalent for that of ${\bf B}_3 $. More precisely
after some rearrangements we find that
\begin{eqnarray}\label{eq:154.5}
\fl {\cal  B}_1'(t) = 2\tilde{{\cal B}}_1(t), \qquad
{\cal  B}_2'(t) = 4\tilde{{\cal B}}_2(t), \qquad
{\cal  B}_3'(t) = {\cal  B}_4'(t) = 2^{3/2}\tilde{{\cal B}}_3(t).
\end{eqnarray}
Comparing (\ref{eq:154.5}) with (\ref{eq:151.3'}) we see that
\begin{equation}\label{eq:155.6}
q_k'(t) = \tilde{q} _k(t) + \ln 2, \qquad k=1,2,3; \qquad q'_4(t)=0,
\end{equation}
where
\begin{equation}\label{eq:155.6'}
\fl \tilde{q}_1(t) = \ln \tilde{{\cal  B}}_1(t), \qquad
\tilde{q}_2(t) = \ln {\tilde{{\cal  B}}_2(t)\over \tilde{{\cal  B}}_1(t)},
\qquad
\tilde{q}_3(t) = \ln {\tilde{{\cal  B}}_3^2(t)\over \tilde{{\cal
B}}_2(t)}.
\end{equation}

The condition (\ref{eq:154.2}), or equivalently, $q_4'(t)=0 $ when imposed
onto the system of equations (\ref{eq:151.3}) leads to a system for
$q'_k(t) $, $k=1$, $2$, $3 $ slightly different from the ${\bf B}_3 $-CTC.
The difference is in the coefficient in front of
$\alpha_3\rme^{-(\vec{q}', \alpha _3)} $ which comes out with an
additional factor of 2. This factor is precisely cancelled if we go over
to the variables $\tilde{q}_k(t) $.

Quite analogously but with more technicalities one can prove that the
symmetry with respect to the outer automorphism $v_2 $ of ${\bf D}_r $
will reduce the ${\bf D}_r $-solution to the one for ${\bf B}_{r-1} $;
more precisely, using analogous notations as above, we can write:
\begin{equation}\label{eq:D-Br}
\fl {\cal  B}_k'(t) = 2^k\tilde{{\cal  B}}_k(t), \quad \mbox{for\;} k=1,
\dots, r-2; \qquad {\cal  B}_{r-1}'(t) ={\cal  B}_r'(t) =
2^{(r-1)/2}\tilde{{\cal B}}_{r-1}(t),
\end{equation}
and
\begin{eqnarray}\label{eq:D-Brq}
q_k'(t) = \tilde{q}_k(t) +\ln 2, \quad \mbox{for\;} k=1,\dots, r-1, \qquad
q_{r}'(t)=0,\nonumber\\
\fl \tilde{q}_k(t) =\ln {\tilde{{\cal B}}_{k}(t) \over \tilde{{\cal
B}}_{k-1}(t)  }, \quad \mbox{for\;} k=1,\dots, r-2,\qquad
\tilde{q}_{r-1}(t) =\ln {\tilde{{\cal B}}_{r-1}^2(t) \over \tilde{{\cal
B}}_{r-2}(t)  }.
\end{eqnarray}
In deriving (\ref{eq:154.5}) we see, that due to $\zeta _4=0 $ two of the
terms in ${\cal  B}_1(t) $ combine together; this corresponds to the fact
that $\Gamma _{{\bf B}_3}(\omega _1) $ has only 7 weights while
$\Gamma _{{\bf D}_4}(\omega _1) $ has 8. Analogous, but more complicated
combinations and cancellations take place in the proof of
(\ref{eq:D-Brq}).
\end{example}

\begin{example}\label{exa:G2}
The case with ${\bf G}_2 $ can be obtained from ${\bf D}_4 $ after
imposing a symmetry condition with respect to the outer automorphism
$v_1$ (\ref{eq:D4-V}) of ${\bf D}_4 $ of order 3. The restrictions that
$v_1$ imposes on the scattering data are:
\begin{eqnarray}\label{eq:V-fix}
\fl \zeta _4=0, \qquad \zeta _1-\zeta _2=\zeta _3, \qquad
\varphi _{04}=0, \qquad \varphi _{01}- \varphi _{02} = \varphi _{03}.
\end{eqnarray}
or in other words
\[
v_1(\vec{\zeta })=\vec\zeta, \qquad v_1 (\vec{\varphi }_0) = \vec{\varphi
}_0.
\]
Then one can check that due to (\ref{eq:149.9}) $\vec{q}(t) $ is
invariant with respect to $v_1 $: $v_1(\vec{q}(t))=\vec{q}(t) $ and,
consequently, is an element of the subalgebra ${\bf G}_2 $.

Indeed, if we average the set of simple roots of ${\bf D}_4 $ with respect
to the action of $v_1 $ and also the system of equations (\ref{eq:151.3})
then we get the system of simple roots of ${\bf G}_2 $:
\begin{equation}\label{eq:G2-rs}
\fl \beta _1 = {1 \over 3 } (\alpha _1+\alpha _3+\alpha _4) = {1\over 3 }
(e_1-e_2+2e_3), \qquad \beta _2 =\alpha _2=e_2-e_3,
\end{equation}
and the system of equations
\begin{equation}\label{eq:152.2}
Q_{1,tt} = \rme^{-(\vec{Q},\beta _1)}, \qquad Q_{2,tt} =
\rme^{-(\vec{Q},\beta _2)} -\rme^{-(\vec{Q},\beta _1)}.
\end{equation}
which is slightly different from (\ref{eq:gCTC}) for ${\bf G}_2 $; namely
the variables, let say $q_1'$, $q_2'$ and $q_3'=-q_1'-q_2'$, entering in
the original system (\ref{eq:gCTC}) for ${\bf G}_2 $ are
\begin{equation}\label{eq:kjynxnaj}
q'_{1}(t) = Q_1(t)-2\ln 3 \qquad q_2'=Q_2-\ln3
\end{equation}
The solution to (\ref{eq:152.2}) is provided by:
\begin{equation}\label{eq:G2-sol}
Q_1=\ln {\cal  B}_1(t;\vec{\zeta }, \vec{\varphi _0}), \qquad
Q_2 = \ln { {\cal  B}_2(t;\vec{\zeta }, \vec{\varphi _0}) \over {\cal
B}_1(t;\vec{\zeta }, \vec{\varphi _0}) }
\end{equation}
where ${\cal  B}_{1,2}(t) $ are obtained from (\ref{eq:149.1}),
(\ref{eq:149.2}) with the additional restrictions (\ref{eq:V-fix});
namely we have:
\begin{eqnarray}\label{eq:G2-Bk}
\fl {\cal  B}_1(t;\vec{\zeta }, \vec{\varphi _0})= 2b_0b_1 \left[
{ \zeta _2-\zeta _3 \over \zeta _1 } \cosh \varphi _1 +
{ \zeta _1 +\zeta _3 \over \zeta _2 } \cosh \varphi _2 +
{ \zeta _1 +\zeta _2 \over \zeta _3 } \cosh \varphi _3 \right] + 2b_1^2,
\nonumber\\
\fl {\cal  B}_2(t;\vec{\zeta }, \vec{\varphi _0}) = b_0b_1^2 \left[
{\zeta _1^2  \over \zeta _2\zeta _3 } { \cosh (\vec{01},\vec{\varphi})
\over \zeta _2-\zeta _3} +
{\zeta _2^2  \over \zeta _1\zeta _3 } { \cosh (\vec{31},\vec{\varphi})
\over \zeta _1+\zeta _3} + {\zeta _3^2  \over \zeta _1\zeta _2 } { \cosh
(\vec{32},\vec{\varphi})\over \zeta _1+\zeta _2} \right. \nonumber\\
\fl \left. + {3(\zeta _1+\zeta_2)\over\zeta _1\zeta _2 }\cosh
(\vec{10},\vec{\varphi}) + {3(\zeta_1+\zeta _3)\over\zeta _1\zeta _3}
\cosh (\vec{11},\vec{\varphi}) + {3(\zeta_2-\zeta _3)\over\zeta _2\zeta _3}
\cosh (\vec{21},\vec{\varphi}) \right] \\
+ 12 b_0^2b_1 \left( {1\over \zeta _3} +{1\over \zeta _2}  - {1\over \zeta
_1}\right), \nonumber\\
b_0 ={1\over 8(\zeta _1 +\zeta _2 )(\zeta _1 +\zeta _3) (\zeta _2 -\zeta
_3 ) }, \qquad  b_1 ={1\over 8\zeta _1 \zeta _2 \zeta _3}, \nonumber
\end{eqnarray}
where by $\vec{ij} $ we denote the root $i\beta _1+j\beta _2 $ of ${\bf
G}_2 $. Obviously each of the terms in ${\cal  B}_k(t) $ in
(\ref{eq:G2-Bk}) can be related to a weight of the corresponding
fundamental representation $\Gamma (\omega _k) $ of ${\bf G}_2 $; so
again we cast the solution in the form (\ref{eq:**a}), (\ref{eq:**b}).
\end{example}

We summarize the results of this section by the following remark.
One can view the solutions of the CTC related to the classical series
${\bf B}_r$, ${\bf C}_r$ as obtained from the Moser formulae combined with
the corresponding constraint on the scattering data. The case of ${\bf
D}_r$ can be obtained likewise if we take the Lax matrix to be
pentadiagonal as in (\ref{eq:ap-3.2}).

On the other hand starting from a Lax matrix related to each of this
series we can always apply Moser's approach and derive as a result the
functions $q_k(t) $, $k=1,\dots, N $. In order that both answers for
the CTC-solutions be compatible one needs to show that
\begin{equation}\label{eq:q-kb}
q_k(t)=-q_{\bar{k}}(t), \qquad \bar{k}=N+1-k
\end{equation}
where $N $ is the dimension of the typical representation of ${\frak  g}$.
Equation (\ref{eq:q-kb}) can be derived from the results in Sections~2 and
3 and shows the compatibility of the two approaches to the CTC solutions.

\section{Dynamical regimes and large time asymptotics}

There are important differences between the RTC and CTC, especially the
asymptotic behavior of their solutions. Indeed, for the RTC, one has
\cite{Moser,Toda} that, both the eigenvalues, $\zeta _k $, and the
constants $\varphi_k(0) $, are always real-valued. Moreover, one can prove
that $\zeta _k \neq \zeta _j $ for $k\neq j $, i.e. no two eigenvalues can
be exactly the same.  As a direct consequence of this, it follows that the
only possible asymptotic behavior in the RTC is an asymptotically
separating, free motion of the particles.

The situation is different for the CTC. Now the eigenvalues $\zeta
_k=\kappa _k + i\eta_k$, as well as the constants $\varphi_k(0)$ become
complex.  Furthermore, the argument of Moser \cite{Moser} does not apply
to the complex case, so one can have multiple eigenvalues.  The collection
of eigenvalues, $\zeta _k$, still determines the asymptotic behavior of
the solutions. In particular, it is $\kappa _k $ that determines the
asymptotic velocity of the $k $-th particle.  For simplicity, we assume
$\zeta _k\neq \zeta_j$ for $k\neq j $.  However, this condition does not
necessarily mean that $\kappa _k \neq \kappa _j $.  We also assume that
the $\kappa _k$'s are ordered as:
\begin{equation}\label{eq:kapp}
\kappa _1 \leq \kappa_2 \leq \dots \leq \kappa _N.
\end{equation}
This ordering is known as the sorting condition. More generally it can
be understood as $-\vec{\kappa } \in \bar{W}_D $ - the closure of the
dominant Weyl chamber. Once this is done, for the corresponding set
of $N$ particles there are three possible general configurations:

A) free particle propagation (Moser case) - then $q_k(t) $ have linear in
$t $ asymptotic behavior and $-\vec{\kappa } $ is in the interior of $W_D
$.

B) bound state(s) and mixed regimes when one (or several) group(s) of
particles form a bound state - then each group of particles oscillate
around common trajectory with linear in $t $ asymptotic behavior; then
$-(\vec{\kappa },\alpha _k)=0 $ for some set of indices $k\in I_{\rm bs}
$.

C)~degenerate solutions when two (or more) of the
eigenvalues $\zeta _k=\zeta _{k+1}=\dots $ are equal - then
$q_k(t)-q_{k+1}(t) $ have logarithmic in $t $ asymptotic behavior, i.e.
the distance between the particles grows as $\ln t$.

Obviously the cases B) and C) have no analogues in the RTC and physically
are qualitatively different from A).

\subsection{Asymptotically free regimes}
 We begin with the first possibility from the above mentioned - free
particle asymptotics (Moser case). It is realized if we require that all
real parts of the eigenvalues $\kappa_k$ are pairwise different; i. e.
$-\vec{\kappa } $ belongs to the interior of the dominant Weyl chamber
$W_D $:
\begin{equation}\label{eq:wech}
(-\vec\kappa,\alpha _s)>0,\qquad s=1,\dots, r,
\end{equation}
while the imaginary parts $\eta_k$ may be arbitrary.

Let us now consider the asymptotics for the ${\bf A}_r $, ${\bf B}_r $,
${\bf C}_r $ and ${\bf D}_r $ series. Using the explicit expressions for
${\cal  B}_k(t) $ it is not difficult to evaluate their asymptotic
behavior for $t\to \pm \infty $:
\begin{equation}\label{eq:Bkas}
{\cal  B}_{k,as}^\pm (t)=W^{(k)}(\vec\zeta,\omega_k^{\pm}
)\rme^{(-2\vec\zeta t+\vec\varphi_0,\omega_k^{\pm})} \left( 1 + {\cal  O}
(\exp(\mp K^\pm_kt))\right).
\end{equation}
Here $\omega_k^{\pm}$ are the highest (lowest) weights, related through
the Weyl group element $w_0$: $\omega ^+_k=w_0(\omega ^-_k)$ and
\begin{eqnarray}\label{eq:5.1}
&& K^+_k = \mathop{\min}\limits_{\gamma \in \Gamma (\omega _k^+)\backslash
\omega _k^+} \re (\vec{\zeta },\gamma  -\omega _k^+ ) = -(\vec{\kappa
},\alpha _k), \nonumber\\
&& K^-_k = \mathop{\min}\limits_{\gamma  \in \Gamma (\omega _k^+)\backslash
\omega _k^-} \re (\vec{\zeta }, \gamma - \omega _k^-) = -(\vec{\kappa },
\alpha _{\tilde{k}}),
\end{eqnarray}
see Appendix~B.

Note the natural way in which the two asymptotics are related by the $w_0
$ transformation of the Weyl group, namely:
\begin{eqnarray}\label{eq:Bk-asw0}
{\cal  B}^-_{k,\rm as} (t) &=& w_0 \left( {\cal  B}^+_{k,\rm as} (t)
\right) = W^{(k)} (\vec{\zeta },w_0(\omega _k^+))
\rme ^{(\vec{\varphi} (t),w_0(\omega _k^+)) } \nonumber\\
&=& W^{(k)} (w_0(\vec{\zeta }),\omega _k^+) \rme ^{(w_0(\vec{\varphi}
(t)),\omega _k^+)}.
\end{eqnarray}
More generally since $W^{(k)}(\vec{\zeta },\gamma) $ depends only on the
scalar products of the type $(\vec{\zeta },\gamma ) $ the action of the
automorphism $w_0 $ on $W^{(k)}(\vec{\zeta },\gamma) $ is given by
$W^{(k)}(\vec{\zeta },w_0(\gamma))=W^{(k)}(w_0(\vec{\zeta }),\gamma) $.

The relations (\ref{eq:Bkas}) and (\ref{eq:5.1}) are due to the simple
fact that the leading exponent for $t\to \infty  $ ($t\to -\infty  $) in $
{\cal B}_k(t) $ corresponds to the weight $\gamma \in \Gamma (\omega _k) $
for which the value of $\re (-\vec{\zeta },\gamma ) $ is maximal
(minimal). Since $-\vec{\kappa }\in W_D $ this maximum (minimum) is
realized when $\gamma =\omega _k^+ $ (respectively, $\gamma =\omega
_k^-$).

From the previous considerations we have:
\begin{equation}\label{eq:5.2}
q_k(t) = \ln {B_k(t)  \over B_{k-1}(t) } = \sum_{s=1}^{r}
{2(\alpha _s, e_k)  \over (\alpha _s, \alpha _s) } \ln {\cal  B}_s(t).
\end{equation}
and consequently the asymptotics $\vec{q}^\pm_{\rm as} $ of $\vec{q}(t) $
for $t\to\pm\infty  $ are given by
\begin{eqnarray}\label{eq:qp-vec}
\fl \vec{q}^+_{\rm as} (t) = -2\vec{\zeta }t +\vec{\varphi _0} +
\sum_{k=1}^{r} {2\alpha _k  \over (\alpha _k,\alpha _k) } \ln W^{(k)}
(\vec{\zeta },\omega _k^+), \\
\label{eq:qm-vec}
\fl \vec{q}^-_{\rm as} (t) = w_0 \left(-2\vec{\zeta }t +
\vec{\varphi _0}\right) + \sum_{k=1}^{r} {2\alpha _k  \over (\alpha
_k,\alpha _k) } \ln W^{(k)} (w_0(\vec{\zeta }),\omega _k^+),
\end{eqnarray}
up to terms falling off exponentially for $t\to\pm \infty  $.
The explicit expressions for the components $q_k(t) $ for the ${\bf A}_r $
series are well known, see e.g., \cite{Moser,Toda,GEI}. For the other
classical series of Lie algebras we obtain:
\begin{eqnarray}\label{eq:qk-asb}
q^\pm _{k,\rm as} (t) = \mp 2\zeta _kt \pm \varphi _{0k}+ \beta _k
+ {\cal  O}\left( \rme ^{\mp N_k^\pm t }\right),
\qquad k=1,\dots,r,
\end{eqnarray}
with
\begin{eqnarray}\label{eq:qk-betak}
\beta _k = \ln \left( w_k \prod_{s=1}^{k-1} (2\zeta _s-\zeta_k)^2\right),
\qquad N_k^\pm = \mathop{\min}\limits_{s: (e_k,\alpha _s)\neq 0} K_s^\pm.
\end{eqnarray}

The only exception to (\ref{eq:qk-asb}), (\ref{eq:qk-betak}) is for
${\frak  g}\simeq {\bf D}_r $ with odd $r $, $k=r $ and $t\to -\infty  $;
then
\begin{equation}\label{eq:betak-ex}
q^- _{r,\rm as} (t) = -2\zeta _rt +\varphi _{0r} +  \ln \left( w_r
\prod_{s=1}^{r-1} (2\zeta _r +2\zeta _s)^2\right).
\end{equation}

It has been known for a long time \cite{Goodman,OPRS} that for the RTC
the asymptotic velocities $\vec{v}^\pm $ are related by
$\vec{v}^-=w_0(\vec{v}^+) $. In the complex case the analogs of $v^\pm $
are the complex vectors $-2\vec{\zeta } $ and $-2w_0(\vec{\zeta }) $
respectively.

Up to now we know of only one physical application of CTC as
a model describing the $N $-soliton train interactions
\cite{GKUE,GUED,Arnold,GEI}.  Getting insight from it we will interprete
$\re q_k(t) $ as the trajectory of the center of mass of the $ k$-th
`particle' (soliton). Besides each particle is complex and possesses an
internal degree of freedom. Then $-2\re \zeta _k=-2\kappa _k $ will be the
asymptotic velocity of the $k $-th particle at $t\to \infty $, while
$-2\im \zeta _k=-2\eta_k $ determines its asymptotic phase velocity.

\subsection{Mixed regimes for ${\bf B}_{r} $, ${\bf C}_{r} $ and ${\bf
D}_{2n} $ }.

Our aim in this and next subsection will be to consider the cases when
two or more particles form bound state(s); we will say that several
particles form a bound state if they have equal asymptotic velocities.
In this subsection we consider only those members of the classical series
for which $w_0\equiv -\mbox{id} $.

Bound state(s) are possible when $-\vec\kappa $ is on the boundaries of
$W_D$, i.e.  if we have
\begin{equation}\label{eq:eqprod}
-(\alpha _k,\vec\kappa )=0, \qquad k\in I_{\rm bs},
\end{equation}
where $I_{\rm bs} \subset \{1,\dots,r\} $ is a subset of indices. If
$I_{\rm bs} =\{m\} $ contains just one index $m<r $ then
$\kappa_{m}=\kappa _{m+1} $ and we will have a two particle bound
state. If $I_{\rm bs} =\{m, m+1,\dots,m+p\} $ and $m+p<r $
then $\kappa _{m} =\kappa _{m+1}= \dots =\kappa _{m+p} $ and we have a
$p+1$-particle bound state. The cases when the largest index in $I_{\rm
bs} $ is equal to $r $ should be considered separately; indeed due to the
fact that the sets of simple roots for the different classical series
differ only in the choices for $\alpha _r $ this cases may lead to
substantially different results.

In our previous paper \cite{GEI} we obtained the large time asymptotics
for the two-particle bound states in the ${\bf A}_r $ CTC. Here we will
briefly analyze more general cases when: a)~${\frak  g} $ belongs to the
other classical series and one bound state may be present, i.e. when
$I_{\rm bs}=\{m\} $, $m\leq r $ and $I_{\rm bs}=\{m,m+1\} $, $m+1\leq r$;
b)~two bound states may be present, i.e. $I_{\rm bs}=\{m, p\} $, $m+1 <p
\leq r $. For brevity we will write down the asymptotics only for those
components $q_k(t) $ which differ from the typical ones (\ref{eq:qk-asb}).
We will limit ourselves with the cases when the mixed regime contains two-
and three-particle bound states only.  The other more complicated regimes
can be analyzed analogously.

Indeed, if for $k\in I_{\rm bs} $ we have $(\vec{\kappa },\alpha _k)=0 $
then at least two terms in ${\cal  B}_k(t) $ may have the same asymptotic
behavior. To our purpose it is sufficient to evaluate only the leading
exponents. Thus we find
\begin{equation}\label{eq:137.2}
\fl {\cal  B}^\pm_{p,\rm as}(t) = \rme^{(\varphi (t),\omega _p^\pm)} \left[
W^{(p)}(\vec{\zeta },\omega _p^\pm) + \sum_{\alpha \in G_p^\pm
(\vec{\kappa })} \bar{W}^{(p)}(\alpha ) + {\cal  O}\left(\rme ^{\mp
K_p^{\prime,\pm} t}\right)\right],
\end{equation}
where
\begin{eqnarray}\label{eq:139.4}
\fl K_p^{\prime,\pm} = \mathop{\min}\limits_{\gamma \in \Gamma
_{p,\pm}(\omega _p^+) } \left[ - 2 (\vec{\kappa }, \omega _p^+ -\gamma
)\right], \qquad
\bar{W}^{(p)}(\alpha ) = \rme ^{-(\varphi (t),\alpha )} W^{(p)}
(\vec{\zeta }, \omega _p^\pm \mp \alpha ), \nonumber\\
G_p^\pm(\vec\kappa ) = \left\{ \alpha >0,\; (\alpha,\vec{\kappa })=0, \;
\pm {2(\alpha,\omega _p^\pm) \over (\alpha,\alpha ) } \geq 1
\right\}, \\
\Gamma _{p,\pm} (\omega _p^+)  = \Gamma (\omega _p^+)
\backslash \left\{ \omega _p^\pm, \omega _p^\pm \mp\alpha, \alpha \in
G_p^\pm(\vec{\kappa }) \right\}. \nonumber
\end{eqnarray}

The condition $(\alpha,\vec{\kappa })=0 $ ensures that $\bar{W}^{(p)} $
only oscillates when $t\to \pm \infty  $ while the third condition in
$G_p^\pm(\vec{\kappa }) $ means that $\omega _p^\pm \mp\alpha \in \Gamma
(\omega _p^+) $.

We start with the simplest case $I_{\rm bs}=\{ m\} $, which contains
several qualitatively different subcases which will be listed below. In
each of them it is possible to describe the sets of roots $G_p^\pm
(\vec{\kappa }) $ and to evaluate the estimating exponents $K_p^{\prime
\pm} $, for details see Appendix~B. It turns out that $G_p^\pm
(\vec{\kappa }) \equiv \emptyset $ for $p\neq m $ and $G_m^+(\vec{\kappa
}) \equiv \{\alpha_m\} $, $G_m^- (\vec{\kappa }) \equiv
G_m^+(w_0(\vec{\kappa })) \equiv \{w_0(\alpha_m)\} $. In what follows we
will concentrate mainly on the asymptotics for $t\to\infty $; the
asymptotics for $t\to -\infty  $ we will obtain by formula
(\ref{eq:qm-vec}).  Therefore the asymptotics of ${\cal B}^+_{p,\rm
as}(t) $ for $p\neq m $ will be given by (\ref{eq:Bkas}) while for ${\cal
B}^+_{m,\rm as}(t)  $ we get:
\begin{equation}\label{eq:139.3}
\fl {\cal  B}^+_{m,\rm as}(t) =\rme ^{(\varphi (t),\omega _m^+)} \left[
W^{(m)}(\vec{\zeta },\omega _m^+) + \bar{W}^{(m)}(\alpha _m) + {\cal  O}
\left( \rme ^{-K_m^{\prime +}t} \right) \right],
\end{equation}
with the following result for $K_m^{\prime +} $ valid for any of the
algebras in the classical series (see Appendix~B):
\begin{equation}\label{eq:139.1}
K_m^{\prime +} =\mathop{\min}\limits_{s: (\alpha _s,\alpha _m)\neq 0}
\left[-2(\vec{\kappa },\alpha _s)\right].
\end{equation}
In other word the minimum should be taken with respect to the simple roots
$\alpha _s $ that are connected to $\alpha _m $ in the Dynkin diagram.

Now we are in position to compare the asymptotic velocities of the
particles and to single out the structure of the bound states (if any).
Since the asymptotic velocity of $q_k(t) $ for $t\to \infty  $ is equal
to $-2\kappa _k $ we just have to see what constraints on $\{ \kappa _1,
\dots, \kappa _r\} $ will be imposed by $-(\vec{\kappa },\alpha _k)>0 $, $
k\neq m $ and $-(\vec{\kappa },\alpha _m)=0 $. For ${\frak  g}\simeq
{\bf B}_r$, ${\bf C}_r$  and $m<r $ we have
\begin{equation}\label{eq:139.6}
\kappa _1 < \dots < \kappa _m =\kappa _{m+1} < \dots <\kappa _r <0;
\end{equation}
for  ${\frak  g} \simeq {\bf D}_{2n}$ and $m<2n-1 $, $m=2n-1 $ we have
\begin{equation}\label{eq:139.7}
\kappa _1 < \dots < \kappa _m =\kappa _{m+1} < \dots <\kappa _{2n-1}
< -|\kappa _{2n}| <0,
\end{equation}
and
\begin{equation}\label{eq:139.8}
\kappa _1 < \dots < \kappa _m <\kappa _{m+1} < \dots <\kappa _{2n-1}
=\kappa _{2n} <0,
\end{equation}
respectively.

{}Finally for ${\frak  g} \simeq {\bf B}_r$, ${\bf C}_r $ and $m=r $ we get:
\begin{equation}\label{eq:139.10}
\kappa _1 < \dots <  \kappa _{r-1} <\kappa _r =0,
\end{equation}
and for ${\frak  g} \simeq {\bf D}_{2n}$ and $m=2n $
\begin{equation}\label{eq:139.9}
\kappa _1 < \dots < \kappa _{2n-1} =-\kappa _{2n} <0.
\end{equation}

{}From (\ref{eq:139.6})-(\ref{eq:139.8}) it is easy to see that for $m<r $
we always have one bound state of two particles ($m $-th and $m+1 $-st);
the rest of the particles go into a free asymptotic regime.
If in addition $w_0=-\mbox{id}\, $, as we assumed in the beginning of this
subsection, this bound state will be present also for $t\to -\infty  $.
Therefore for this classes of algebras we have stable 2-particle bound
states for all $m<r $.

{}For $m=r $ the situation is different. From (\ref{eq:139.10}) we see that
the condition $-(\vec{\kappa },\alpha _r)=0 $ for ${\frak  g} \simeq {\bf
B}_r$, ${\bf C}_r $ just means that the $r $-th particle has vanishing
velocity. As for ${\frak  g} \simeq {\bf D}_{2n}$ the condition
$-(\vec{\kappa },\alpha _{2n})=0 $ means that the $2n-1 $-st and the $2n
$-th particle have opposite velocities. Therefore for $m=2n $ no bound
states are possible.

The next possibility is that the set $I_{\rm bs} \equiv \{m,p\}$. There
are qualitatively different cases here: a)~$(\alpha _m,\alpha _p)=0 $ and
b)~$(\alpha _m,\alpha _p)\neq 0 $. Each of the values $m $ and $p $ in
case a) can be considered independently and to each of them applies the
analysis already posed above. In the generic case $m<p<r $ we will have
two pairs of bound states each containing two particles; if $m<p=r $ then
we have only one bound state of two particles. An exception here is the
case ${\frak  g}\simeq {\bf D}_{2n} $ and $m=2n-1 $, $p=2n $. The two
roots $\alpha _{2n-1} $ and $\alpha _{2n} $ are obviously orthogonal, but
now the condition (\ref{eq:eqprod}) leads to $\kappa _{2n}=\kappa
_{2n-1}=0 $ and as s result in this case we have only one bound state
consisting of two particles with vanishing velocities.

Let us now analyze the case b). For generic values of $m<p<r $ what we
find is a bound state of three particles. One possible realization of b)
is to take $p=m+1<r $; then the condition (\ref{eq:eqprod}) leads to
\begin{equation}\label{eq:139.III}
\kappa _1<\dots <\kappa _m =\kappa _{m+1} =\kappa _{m+2} <\dots
\kappa_{m+3} <\dots,
\end{equation}
i.e., the particles numbered by $m $, $m+1 $ and $m+2 $ move with the same
asymptotic velocities and form a bound state. Again we must look through
all possibilities when case b) takes place and point out possible
exceptions. Such for example is the case when ${\frak  g}\simeq {\bf B}_r
$, ${\bf C}_r $ and $m=r-1 $, $p=r $. Equation (\ref{eq:eqprod}) then
gives $\kappa _{r-1}=\kappa _{r}=0 $, which means that this is bound state
of two particles: $r-1 $-st and the $r $-th with vanishing velocity.

If ${\frak  g}\simeq {\bf D}_{2n}$  and $m=2n-2 $, $p=2n-1 $ we get a
three-particle bound state with velocity $\kappa _{2n-2}=\kappa
_{2n-1}=\kappa _{2n}<0 $. The last example related to this algebra is
$m=2n-2 $, $p=2n $ which corresponds to $\kappa _{2n-2}=\kappa
_{2n-1}=-\kappa _{2n} $.  This means that the particles $2n-2 $ and $2n-1
$ form a bound state, but the last $2n $-th particle moves with the
opposite velocity and is not a part of the bound state.

Obviously the number of examples can be extended to include sets  $I_{\rm
bs} $ with more indices; one can expect to have bound states with
increasing number of bounded particles. It is not difficult to present
also the explicit form of the asymptotics of $q_{k,\rm as}^\pm (t) $. The
most difficult part in this calculation is to determine the sets of roots
$G_p^\pm(\vec\kappa ) $. We list these sets of roots in Appendix~B for the
classical series of Lie algebras related to the sets $I_{\rm bs} $ with
one and two indices. Indeed if choose ${\frak  g} \simeq {\bf B}_r$,
$I_{\rm bs}=\{ r-1,r\} $. Then the sets of roots
\[
\fl G_{r-1}^+(\vec\kappa )=\{\alpha _{r-1},  \alpha _{r-1}+\alpha _{r},
\alpha_{r-1} +2\alpha _{r}\}, \qquad G_{r}^+(\vec\kappa )=\{\alpha _{r},
\alpha _{r-1}+\alpha_{r}, \alpha_{r-1} +2\alpha _{r}\},
\]
Then
\begin{eqnarray}\label{eq:140.2}
\fl {\cal  B}_{r-1,\rm as}^+ = \rme ^{(\varphi (t),\omega ^+_{r-1})} \left[
W^{(r-1)}(\vec{\zeta },\omega _{r-1}^+) + \bar{W}^{(r-1)}(\alpha _{r-1})
\right. \nonumber\\ +
\left. \bar{W}^{(r-1)}(\alpha _{r-1}+ \alpha _r) +
\bar{W}^{(r-1)}(\alpha _{r-1}+ 2\alpha _r) \right], \\
\fl {\cal  B}_{r,\rm as}^+ = \rme ^{(\varphi (t),\omega ^+_{r})} \left[
W^{(r)}(\vec{\zeta },\omega _{r}^+) + \bar{W}^{(r)}(\alpha _{r}) +
\bar{W}^{(r)}(\alpha _{r-1}+ \alpha _r) +  \bar{W}^{(r)}(\alpha _{r-1}+
2\alpha _r) \right].\nonumber
\end{eqnarray}
Now we have to insert (\ref{eq:140.2}) into (\ref{eq:5.2}). After some
calculations we obtain $\kappa _{r-1}=\kappa _{r}=0 $ and
\begin{eqnarray}\label{eq:140-2}
 q^+_{r-1,\rm as} (t) = - 2i\eta_{r-1} t + \varphi _{0r-1} + \ln {
\Sigma_{r-1} \over W^{(r-2)}(\vec{\zeta },\omega _{r-2}^+)}, \nonumber\\
 q^+_{r,\rm as} (t) = - 2i\eta_{r} t + \varphi _{0r} + \ln {\Sigma_{r}^2
\over \Sigma _{r-1}}, \\
\Sigma _p = W^{(p)}(\vec{\zeta },\omega _p^+) + \tilde{\Sigma }_p,
\qquad p=r-1,r, \nonumber\\
\tilde{\Sigma }_p = \bar{W}^{(p)}(\alpha _p) + \bar{W}^{(p)}(\alpha
_{r-1}+\alpha _r)+ \bar{W}^{(p)}(\alpha _{r-1}+2\alpha _r).\nonumber
\end{eqnarray}

In this subsection we presented various  types of mixed regimes which
could be called regular. By regular here we mean that the number and the
structure of the bound states at $t\to -\infty  $ coincides with the ones
for $t\to \infty  $. In the next subsection we consider the `irregular'
mixed regimes, which change qualitatively their structure during the
evolution.

\subsection{Mixed regimes for ${\bf D}_{2n+1} $: `creation' and
`decay' of bound states.  }

The `irregular' mixed regimes take place only for algebras for which
$w_0\neq -\mbox{id} $. This takes place for ${\frak  g}\simeq {\bf A}_r $
and ${\frak  g}\simeq {\bf D}_{2n+1} $. At the end of this subsection we
will explain why the `irregular' regimes, i.e., the effects  of `creation'
and `decay' of bound states can be related only to ${\frak  g}\simeq {\bf
D}_{2n+1} $.

{}First of all we note that most of the bound states related to ${\bf
D}_{2n+1} $ are regular. Such are the states corresponding to $I_{\rm
bs}=\{m\} $ with $m<2n $ and $I_{\rm bs}=\{m,p\} $ with $m<p<2n $.
The formulas for the asymptotics in all these cases are quite analogous to
the ones already presented.

Let us start with the first `irregular' case with $I_{\rm bs}=\{2n\} $.
This means that $\kappa _{2n}=\kappa _{2n+1} $ and for
$t\to\pm\infty  $
\begin{eqnarray}\label{eq:D-q-r-1}
q^\pm_{2n, \rm as} = \mp 2\kappa _{2n} t \mp 2i\eta_{2n}t \pm \varphi
_{0,2n} + \beta ^{\pm,\prime}_{2n}(\vec{\zeta }), \nonumber\\
q^\pm_{2n+1, \rm as} = - 2\kappa _{2n} t - 2i\eta_{2n+1}t +
\varphi_{0,2n+1} + \beta ^{\pm,\prime}_{2n+1}(\vec{\zeta }), \\
\beta ^{+,\prime}_{2n}(\vec{\zeta })= \ln {\left[ W^{(2n)}(\vec{\zeta
},\omega _{2n}^+) + \bar{W}^{(2n)}(\alpha _{2n}) \right]
W^{(2n+1)}(\vec{\zeta },\omega _{2n+1}^+) \over W^{(2n-1)}(\vec{\zeta
},\omega _{2n-1}^+) }, \nonumber\\
\beta ^{+,\prime}_{2n+1}(\vec{\zeta })=\ln {W^{(2n+1)}(\vec{\zeta },\omega
_{2n+1}^+) \over W^{(2n)}(\vec{\zeta },\omega _{2n}^+) +
\bar{W}^{(2n)}(\alpha _{2n}) }. \nonumber
\end{eqnarray}
and $\beta ^{-,\prime} _{k}(\vec\zeta )$, $k=2n $, $2n+1 $ are obtained
from $\beta ^{+,\prime} _{k}(\vec\zeta )$ by using (\ref{eq:beta-w}).
Obviously at $t\to -\infty $ the $2n$-th and the $2n+1 $-st particles have
opposite velocities, while for $t\to \infty  $ their velocities become
equal. This situation can be viewed as `creation' of a bound state.

The second `irregular' case is with $I_{\rm bs}=\{2n+1\} $.
This means that $\kappa _{2n}=-\kappa _{2n+1} $ and for $t\to\pm\infty  $
\begin{eqnarray}\label{eq:D-q-r}
q^\pm_{2n, \rm as} = \mp 2\kappa _{2n} t \mp 2i\eta_{2n}t \pm \varphi
_{0,2n} + \beta ^{\pm,\prime\prime}_{2n}(\vec{\zeta }), \nonumber\\
q^\pm_{2n+1, \rm as} =  2\kappa _{2n} t - 2i\eta_{2n+1}t +
\varphi_{0,2n+1} + \beta ^{\pm,\prime}_{2n+1}(\vec{\zeta }), \\
\beta ^{+,\prime\prime}_{2n}(\vec{\zeta })= \ln {\left[
W^{(2n+1)}(\vec{\zeta},\omega _{2n+1}^+) + \bar{W}^{(2n+1)}(\alpha _{2n+1})
\right] W^{(2n)}(\vec{\zeta },\omega _{2n}^+) \over
W^{(2n-1)}(\vec{\zeta },\omega_{2n-1}^+) }, \nonumber\\
\beta ^{+,\prime\prime}_{2n+1}(\vec{\zeta })=\ln {W^{(2n+1)}(\vec{\zeta
},\omega _{2n+1}^+) + \bar{W}^{(2n+1)}(\alpha _{2n+1}) \over
W^{(2n)}(\vec{\zeta },\omega_{2n}^+) }. \nonumber
\end{eqnarray}
and again $\beta ^{-,\prime} _{k}(\vec\zeta )$, $k=2n $, $2n+1 $ are
obtained from $\beta ^{+,\prime} _{k}(\vec\zeta )$ by using
(\ref{eq:beta-w}).  Now at $t\to -\infty  $ the $2n$-th and the $2n+1 $-st
particles have equal velocities, while for $t\to \infty  $ their
velocities become opposite. This situation can be viewed as `decay' of a
bound state.

The next more complex situation is when $I_{\rm bs}=\{m,p\} $. Again we
should consider two distinct subcases, namely $(\alpha _p,\alpha _m)=0 $
and $(\alpha _m,\alpha _p)\neq 0 $.

In both cases we recover `regular' asymptotics provided $m<p<2n $;
namely for such choices of $I_{\rm bs} $ we have either two pairs of
two-particle bound states (if $(\alpha _m,\alpha _p)= 0$) or a
three-particle bound state (if $(\alpha _m,\alpha _p)\neq 0 $).

The `irregular' cases with $(\alpha _m,\alpha _p)= 0 $ are two types. The
first one takes place if $m<2n $ and $p=2n $  ($p=2n+1 $). Then for $t\to
-\infty  $ ($t\to \infty $) we have two bound states formed by the
particles $\{m, m+1\} $ and $\{2n,2n+1\} $ while at $t\to \infty  $ ($t\to
-\infty  $) the second bound state decays and we are left with only one
bound state.  Quite different is the situation when $I_{\rm
bs}=\{2n,2n+1\} $. This corresponds to $\kappa _{2n}=\kappa _{2n+1}=0 $,
so this is a regular case but with only {\em one bound state} formed by
the particles $\{2n,2n+1\} $ with vanishing velocity.

There are only two `irregular' cases with $(\alpha _m,\alpha _p)\neq 0 $,
namely $I_{\rm bs}=\{2n-1,2n\} $ and $I_{\rm bs}=\{2n-1,2n+1\} $. The
first one leads to $\kappa _{2n-1}=\kappa _{2n}=\kappa _{2n+1}<0 $ and to
the following asymptotic behavior of $q_k(t) $, $k=2n-1 $, $2n $ and
$2n+1 $:
\begin{eqnarray}\label{eq:D2n+1-3}
q^\pm_{2n-1,\rm as} = \mp 2\kappa _{2n-1}t \mp 2i\eta_{2n-1}t \pm
\varphi _{0,2n-1} + \beta _{2n-1}^{\prime,\pm}, \nonumber\\
q^\pm_{2n,\rm as} = \mp 2\kappa _{2n-1}t \mp 2i\eta_{2n}t \pm
\varphi _{0,2n} + \beta _{2n}^{\prime,\pm}, \\
q^\pm_{2n+1,\rm as} = - 2\kappa _{2n-1}t - 2i\eta_{2n+1}t +
\varphi _{0,2n+1} + \beta _{2n+1}^{\prime,\pm}, \nonumber\\
\fl \beta _{2n-1}^{\prime,+} = \ln {\Sigma _{2n-1}^{\prime,+}(\vec{\zeta })
\over W^{(2n-2)}(\vec{\zeta },\omega ^+_{2n-2}) }, \qquad
\beta _{2n}^{\prime,+} = \ln {\Sigma _{2n}^{\prime,+}(\vec{\zeta })
W^{(2n+1)}(\vec{\zeta },\omega ^+_{2n+1}) \over
\Sigma _{2n-1}^{\prime,+}(\vec{\zeta }) }, \nonumber\\
\beta _{2n+1}^{\prime,+} = \ln { W^{(2n+1)}(\vec{\zeta },
\omega ^+_{2n+1})\over \Sigma _{2n}^{\prime,+}(\vec{\zeta })   },
\nonumber\\
\Sigma _p^{\prime,+}(\vec{\zeta }) = W^{(p)}(\vec{\zeta },\omega
^+_{p}) + \bar{W}^{(p)}(\alpha _p) +\bar{W}^{(p)}(\alpha _{2n-1}
+\alpha _{2n}).\nonumber
\end{eqnarray}

{}From these formulae we find out that for $t\to -\infty  $ we have
a two particle bound state formed by $2n-1 $-st and $2n $-th particles,
while for $t\to \infty $ the $2n+1 $-st particle `joins' them and we
have a three-particle bound state.

The case with $I_{\rm bs}=\{ 2n-1,2n+1\} $ is analogous: the only
difference is that at $t\to -\infty  $ we have a three-particle bound state
formed by the $2n-1 $-st, $2n $-th and $2n+1 $-st particles,
while for $t\to \infty  $ the $2n+1 $-st particle `separates' from them
and we are left with a two-particle bound state.

Let us analyze this situation on the bases of our remark at the end of
Section~3. Let us first explain why such irregular solution are not
possible for the ${\bf A}_r $-series. In this case we have $r+1 $
particles and the sets of asymptotic velocities for $t\to \infty  $ and
$t\to -\infty  $ differ only in the ordering: $\{ -2\kappa _{r+1}\leq
-2\kappa _r \leq \dots \leq -2\kappa _1\} $ and $\{ -2\kappa _{1}\geq
-2\kappa _2 \geq \dots \geq -2\kappa _{r+1}\}  $ respectively. That is why
it was quite natural to identify the $k $-th particle at $t\to -\infty  $
with the $\bar{k} $-th particle at $t\to \infty $: they move with equal
velocities. This is compatible with the action of $w_0 $ in the ${\bf A}_r
$ case, see (\ref{eq:w0a}). As a result if we have, say two bound states
at $t\to -\infty  $, i.e. $-2\kappa _1 =-2\kappa _2 >-2\kappa_3=-2\kappa
_4  $ at $t\to \infty  $ we will have again two bound states $-2\kappa
_{r+1} =-2\kappa _r <-2\kappa_{r-1}=-2\kappa _{r-2}   $.

Next we can view the solutions of CTC related to the classical series
${\bf B}_r $, ${\bf C}_r $ and ${\bf D}_r $ as special symmetric solutions
of the $sl(N) $-CTC, see (\ref{eq:q-kb}).  Then it is enough to consider
only `half' of the trajectories; the other half is obtained  as a `mirror'
image. In this situation the sets of initial and final velocities are
different and the identification, good for ${\bf A}_r $ are not possible
for the other classical series; quite different is also the action of
$w_0$ on the orthonormal basis of the root space, see (\ref{eq:w0a}).

As we mentioned above, if we consider the whole picture with
all the $N $-trajectories we will see that no `creation' or `decay' of
bound states take place. In the cases of ${\bf D}_{2n+1} $ and
$(\vec{\kappa },\alpha _{2n})=0 $ (or $(\vec{\kappa },\alpha _{2n+1})=0 $)
this can be explained as follows. For $t\to \infty $ ($t\to -\infty $) we
have a bound state between the $2n $-th and $2n+1 $-st particle which for
$t\to -\infty  $ ($t\to \infty  $ ) transfers into a bound state between
the $2n $-th and its `mirror' symmetric $2n+2 $-th particle.  `Cutting'
off the symmetric trajectories with numbers $2n+2,\dots,N=4n+2 $ we find
the effects described above.

Analogously we can explain the situation with a three-particle bound state
at $t\to\pm\infty  $ and a two-particle bound state at
$t\to \mp\infty  $. The whole picture of $4n+2 $ particles will always
contain a three-particle bound state.

\subsection{Bound state regimes. Periodic and singular regimes.}

These regimes take place if $\vec{\kappa }=0 $, i.e. the set of
eigenvalues $\zeta _k = i\eta_k $ are purely imaginary. Then each of the
functions $B_k(t) $ will be generically bounded. In particular this means
that all the complex `particles` (or solitons) will move together forming
a bound state with a large number of degrees of freedom. In order to avoid
degeneracies we have to request that $\vec{\eta}\in W_D $.

In order to have periodic solutions we need one more restrictions upon
$\eta_k $, namely
\begin{equation}\label{eq:7.1}
\eta_k -\eta_m = s_{km}\eta_0,
\end{equation}
where $s_{km} $ are integers; if $s_{km} $ are rational we can always make
them integers by rescaling $\eta_0 $.

\begin{example}\label{exa:B2}
Let ${\frak  g}\simeq {\bf B}_2 =so(5) $. The corresponding equations have
the form:
\begin{equation}\label{eq:141.6}
{d^2 q_1 \over dt^2 } = \rme ^{q_2-q_1}, \qquad
{d^2 q_2 \over dt^2 } =- \rme ^{q_2-q_1} + \rme ^{-q_2},
\end{equation}
and their periodic solutions are given by:
\begin{equation}\label{eq:141.7}
q_1 =\ln {\cal  B}_1(t), \qquad
q_{2} =\ln {{\cal  B}_2^2(t) \over {\cal  B}_1(t) },
\end{equation}
where
\numparts
\begin{eqnarray}\label{eq:141a.6a}
\fl {\cal  B}_1(t) &=& {1 \over 16\eta_0^4 (p_1^2 - p_2^2) } \left\{
{\cos 2p_1(\Phi(t) +\Gamma) \over p_1^2 } + {\cos 2p_2(\Phi(t) -\Gamma)
\over p_2^2 } + {p_1^2 - p_2^2  \over p_1^2 p_2^2 } \right\}, \\
\label{eq:141a.6}
\fl {\cal  B}_2(t) &=& {-i  \over 8\eta_0^3 p_1 p_2 } \left\{
{\cos (p_+\Phi(t) + p_-\Gamma) \over p_+ } + {\cos (p_-\Phi(t) + p_+\Gamma)
\over p_- }  \right\},
\end{eqnarray}
\endnumparts
where $\eta_k=p_k\eta_0 $, $p_k $ are integers and
\begin{equation}\label{eq:141a.5}
\fl \Phi (t)=\eta_0 t + {i  \over 4 } \left( {\phi _{01}  \over p_1 } +
{\phi _{02}  \over p_2 } \right), \qquad \Gamma =  {i  \over 4 } \left(
{\phi _{01}  \over p_1 } - {\phi _{02}  \over p_2 } \right), \qquad
p_\pm = p_1\pm p_2.
\end{equation}
The period is provided by
\begin{equation}\label{eq:141a.10}
\tau = {\pi  \over  \eta_0 s_0}
\end{equation}
where $s_0 $ is the greatest common divisor of $p_1 $, $p_2 $, $p_+ $ and
$p_- $.

\end{example}

Our next remark is that in the generic case when $\re \phi _{0k}\neq 0 $
the solution (\ref{eq:141a.6a}), (\ref{eq:141a.6}) is a regular one;
then $|{\cal  B}_1(t)| $ and $|{\cal  B}_2(t)| $ are strictly positive for
all $t $. If we choose however, $\re \phi _{01} =\re \phi _{02}=0  $ then
$|{\cal  B}_1(t)| $ and $|{\cal  B}_2(t)| $ may vanish and the
corresponding  solutions $q_1(t) $ and $q_2(t) $ become singular. Due to
the periodicity, if $|{\cal  B}_1(t)| $ and $|{\cal B}_2(t)| $ vanish at
certain points $t_{01} $ and $t_{02} $ respectively, then they will vanish
also at $t_{01} +k\tau $ and $t_{02} +k\tau $ for any integer $k=0 $, $\pm
1$, $\pm 2 $,\dots.

\begin{example}\label{exa:C2}
Let ${\frak  g}\simeq {\bf C}_2 =sp(4) $. Since the algebras ${\bf B}_2
\simeq {\bf C}_2 $ then the corresponding solutions differ by change of
variables. Let us denote all variables of the ${\bf C}_2 $-CTC model by
the same letters as for ${\bf B}_2 $, adding additional `bar' to
distinguish between them. Then the ${\bf C}_2 $-CTC system has the form:
\begin{equation}\label{eq:142.10}
{d^2 \bar{q}_{1}\over dt^2} = \rme^{\bar{q}_2-\bar{q}_1}, \qquad
{d^2 \bar{q}_{2}\over dt^2} = -\rme^{\bar{q}_2-\bar{q}_1} + 2 \rme
^{-2\bar{q}_2},
\end{equation}
and the solution is presented by:
\begin{equation}\label{eq:143.3}
\bar{q}_1(t) = \ln \bar{{\cal  B}}_1(t) + \ln 2, \qquad
\bar{q}_2(t) = \ln {\bar{{\cal  B}}_2(t) \over \bar{{\cal  B}}_1(t)} + {1
\over 2 } \ln 2,
\end{equation}
where
\begin{equation}\label{eq:143.2}
\fl \bar{{\cal  B}}_1(t) = 2 {\cal  B}_2(t, \bar{\zeta _1}, \bar{\zeta
_2}, \bar{\phi}_{01}, \bar{\phi}_{02}),\qquad \bar{{\cal  B}}_2(t) = 2
{\cal B}_1(t, \bar{\zeta _1}, \bar{\zeta _2}, \bar{\phi}_{01},
\bar{\phi}_{02}),
\end{equation}
and ${\cal  B}_k(t) $ are given by(\ref{eq:141a.6a}), (\ref{eq:141a.6}) and
\begin{eqnarray}\label{eq:143.1}
\bar{\zeta }_1 =\zeta _1 +\zeta _2, \qquad \bar{\zeta }_2 =\zeta _1
-\zeta _2, \nonumber\\ \bar{\phi }_{01} =\phi _{01} + \phi _{02}, \qquad
\bar{\phi }_{02} =\phi _{01} - \phi _{02}.
\end{eqnarray}

Of course these last two examples are analytic continuations of the
solutions presented in \cite{LeSav}.
\end{example}

In analogy to the previous example we may assume $\zeta _k $ to be
purely imaginary with $\eta_k=p_k\eta_{0} $ with integer $p_k $. Then we
get the corresponding periodic solutions to Eq.~(\ref{eq:142.10}). More
generally, inserting purely imaginary values for $\zeta _k $ will result
in periodic solution in all the above examples. These solutions may become
singular if the corresponding parameters $\phi _{0k} $ are purely
imaginary.

\subsection{Degenerate solutions}\label{ssec:}

Let us briefly discuss the degenerate solutions to the CTC. The
degeneracy is possible only if the matrix $L(0) $ has nontrivial Jordan
cells \cite{GEI}.

One possibility  to derive the degenerate solutions is to evaluate the
limit $\zeta _1\to\zeta _2 \to \dots \to \zeta _k $ of the solution
(\ref{eq:24.2}), (\ref{eq:24.3}) using the l'Hospital rule.

If in particular, we have complete degeneracy (i.e., all $\zeta_k$ are
equal to zero) the solution of the $sl(N) $-CTC  can be obtained in a
more simple way. Since all $B_k(t)$ are  polynomials  in $t$ which must
satisfy
\begin{eqnarray}\label{eq:grp}
B_0(t)=B_N(t)=1,\qquad
\ddot{B}_{k}B_{k}-\dot{B}_{k}^{2}=B_{k-1}B_{k+1}.
\end{eqnarray}
we find that they depend on $N-1 $ constants $f_k $, $k=1,\ldots, N-1 $.
More specifically $B_k(t)$ must be polynomial of degree  $k(N-k)$ whose
coefficients can be determined explicitly for example, by the method of
undefined constants.

{}For example, for $N=3$ and $\zeta_1=\zeta _2=\zeta _3=0 $ we get:
\begin{eqnarray}\label{eq:q12}
B_{1}(t)= -{1\over 2}t^{2}+f _{1}t+f _{2}, \qquad
B_{2}(t)= -{1\over 2}t^{2}+f _{1}t -f _{1}^{2}-f _{2},
\end{eqnarray}
and $B_3=1 $, where $f_k $, $k=1,2 $ are complex constants. If $f_1 $ is
real we may use the translational invariance of the CTC-equation and
change $t\to t+f_1 $ to eliminate it; then the solution (\ref{eq:q12}) can
be written in the form:
\begin{equation}\label{eq:q23}
\fl B_{1}(t)= -{1\over 2}t^{2}+F _{1}, \qquad B_{2}(t)= -{1\over 2}t^{2}-
{}F _{1}, \qquad F_1= {1  \over 2 } f_1^2 +f_2.
\end{equation}
If $F_1 $ is real, then the solutions $q_k(t) $ have singularities for
$t=\pm \sqrt{2|F_1|} $. The large time asymptotics are given by:
\begin{equation}\label{eq:as3}
q_{1,\rm as}^\pm(t) = -q_{3,\rm as}^\pm(t) = 2\ln t -\ln 2 +i\pi, \qquad
q_{2,\rm as}^\pm(t) =0,
\end{equation}
i.e., they do not depend on the constants $f_k $ and are complex.
Analogously for $N=4$ and $\zeta_1=\zeta _2=\zeta_3=\zeta_4=0 $ we find:
\begin{eqnarray}\label{eq:01}
\fl B_0(t)=B_4(t)=1,\qquad  B_{1}(t)= {1\over 6}
(t^{3}+f _{1}t^2+f _{2}t+f_3),\nonumber\\
\label{eq:4q22}
\fl B_{2}(t)= -{1\over 12}t^{4}-{1\over 9}f _{1}t^3 -{1\over 18}f
_{1}^{2}t^2-{1\over 18}(f_1f _{2}-3f_3)t-{1\over 36}(f_2^2-2f_1f_3), \\
\label{eq:4q23}
\fl B_{3}(t)= -{1\over 6}t^3 -{1\over 6}f_{1}t^2-({1\over 9} f_1^2-{1\over
6}f_2)t-{1\over 54}(2f_1^3-6f_1f_2+9f_3), \nonumber
\end{eqnarray}
where $f_k $, $k=1,2,3 $ are complex constants. If $f_1 $ is real we
may change $t\to t+f_1/3 $ to eliminate one of these constants; then the
solution (\ref{eq:01}) can be written in the form:
\begin{eqnarray}\label{eq:4r12}
B_1(t)= {1  \over 6 } \left( t^3 + F_2 t + F_3 \right), \qquad
B_2(t)= -{1  \over 36 } \left( 3t^4 - 6 F_3 t + F_2^2 \right), \nonumber\\
B_3(t)= -{1  \over 6 } \left( t^3 - F_2 t + F_3 \right),
\end{eqnarray}
where $F_1 $ and $F_2 $ are expressed through $f_k $ by
\begin{equation}\label{eq:4r15}
{}F_2 = f_2 - {f_1^2  \over 3 }, \qquad
{}F_3 = f_3 + {2 f_1^3  \over 27 } - {f_1f_2  \over 3 }.
\end{equation}

Obviously, these solutions will be regular if $B_k(t) $ have complex roots
and will develop singularities if one (or more) of their roots are real.
More specifically if $F_2=0 $ the solution becomes symmetric, i.e.
$B_1(t)=-B_3(t) $ and has a singularity at $t=0 $. If in addition $F_3 $
is real, then there are singularities also for $t=-\sqrt[3]{F_3} $ and
$t=\sqrt[3]{2F_3} $.

The asymptotics of these solutions are easy to calculate:
\begin{eqnarray}\label{eq:d4as1}
&& q_{1,as\pm}(t)=3\ln t -\ln 6, \qquad q_{2,as\pm}(t)=\ln t -\ln 2 -\rmi
\pi, \\ \label{eq:d4as3}
&& q_{3,as\pm}(t)=-\ln t +\ln 2, \qquad q_{4,as\pm}(t)=-3\ln t +\ln 6 +\rmi
\pi.
\end{eqnarray}
Note that again these asymptotics: a)~do not depend on the constants
$F_k$, and b)~are always complex. The last property is a consequence of
the fact that degeneracy is possible only for the CTC.

\section{Conclusions}\label{}

Detailed analysis of the properties of the fundamental representations of
the simple Lie algebras allowed us to propose an effective and invariant
parametrization for the solutions of the CTC. These solutions describe
much richer asymptotical regimes as compared to the RTC. The explicit
solutions proposed above allow one to evaluate explicitly the large time
asymptotics for the whole variety of dynamical regimes.

{}Further investigation deserve also the degenerate solutions.

One final remark is that one more step is necessary for the perfection of
the explicit formulae (\ref{eq:**a}), (\ref{eq:**b}), namely one should
look for an invariant expression for the functions
$W^{(k)}(\vec{\zeta},\gamma _I^{(k)}) $ in terms of $\gamma ^{(k)}_I $
and the roots system $\Delta  $ only.  Work in this direction is in
progress.

\ack

One of us (VSG) is grateful to Professors  E. Horozov, V.
Kuznetsov and P.  van Moerbecke for useful discussions.

\appendix
\section{The properties of $V $.}

Here we outline some of the details in deriving the expressions for
$A_k(t) $ and $B_k(t) $. As we mentioned in Section~2 we need the explicit
expressions for the minors of $V $.

Let us consider the eigenvalue problem (\ref{eq:eigve}) and let us make
use of the explicit tridiagonal form of $L(0) $. Then it is not difficult
to find that the eigenvector related to $\zeta _k $ is of the form:
\begin{equation}\label{eq:ap-1.1}
v^{(k)}= \left(\begin{array}{c} r_k p_1 \\
r_k p_2 (\zeta _k + P_0(\zeta _k)) \\
r_k p_3 (\zeta _k^2 + P_1(\zeta _k)) \\
\vdots \\
r_k p_{N} (\zeta _k^{N-1} + P_{N-2}(\zeta _k))
\end{array} \right),
\end{equation}
where $P_s(\zeta_k ) $ stands for a polynomial of degree $s $ in $\zeta _k
$ and
\begin{equation}\label{eq:ap-1.2}
\fl
p_1=1,\qquad p_2= { 1 \over  a_1(0)}, \qquad p_3= {1\over a_1(0) a_2(0)},
\qquad p_{N}={ 1 \over  a_1(0)\dots a_{N-1}(0)}.
\end{equation}
Next we note that the terms with $P_s(\zeta _k) $ do not contribute to the
minors $V\left\{ \begin{array}{ccc} 1 & \dots & k\\ i_1 & \dots & i_k
\end{array}\right\} $ and so:
\begin{equation}\label{eq:ap-2.1}
V\left\{ \begin{array}{ccc} 1 & \dots & k\\ i_1 & \dots & i_k
\end{array}\right\}  = r_{i_1} \dots r_{i_k} p_1\dots p_k W(i_1,\dots,
i_k),
\end{equation}
where
\begin{equation}\label{eq:ap-2.2}
\fl W(i_1,\dots, i_k) = \det \left| \begin{array}{ccc} 1 & \dots & 1 \\
2\zeta _{i_1} & \dots & 2\zeta _{i_k} \\
\vdots & & \vdots \\
(2\zeta _{i_1})^{k-1} & \dots & (2\zeta _{i_k})^{k-1}
\end{array} \right| = \prod_{s>p; s,p\in I}^{} 2(\zeta _s - \zeta _p),
\end{equation}
is the Vandermonde determinant and $I=\{ i_1,\dots, i_k\} $. Next we have
to take care of the factors $p_s $, which can be expressed through
$\vec{q}(0) $ since $a_k(0) = \case{1}{2} \exp(-(\vec{q}(0),\alpha _k)/2)
$. We remind also that $r_k $ are determined up to a sign by the
normalization condition (\ref{eq:norm}). These remarks and the properties
of the fundamental representations of the series ${\bf A}_r $
(\ref{eq:Al-4}), (\ref{eq:Al-5}) are sufficient to treat the ${\bf A}_r $
series.

Let us now derive the symmetry relations (\ref{eq:26a}), (\ref{eq:26b})
for the ${\bf B}_r $ and ${\bf C}_r $ algebras. To this end we introduce
the matrices $S $ as follows:
\begin{eqnarray}\label{eq:S-mat}
S &=& \sum_{k=1}^{r} (-1)^{k+1} \left( E_{k\bar{k}} + E_{\bar{k}k}\right)
+  (-1)^{r} E_{r+1,r+1} \qquad \mbox{for\;} {\bf B}_r, \nonumber\\
&=& \sum_{k=1}^{r} (-1)^{k+1} \left( E_{k\bar{k}} - E_{\bar{k}k}\right)
\qquad \mbox{for\;} {\bf C}_r, \\
&=& \sum_{k=1}^{r} (-1)^{k+1} \left( E_{k\bar{k}} + E_{\bar{k}k}\right)
\qquad \mbox{for\;} {\bf D}_r, \nonumber
\end{eqnarray}
which enter into the definition of the corresponding orthogonal and
symplectic algebras. By $E_{jk} $ we denote an $N\times N $ matrix whose
matrix elements are equal to $(E_{jk})_{mn}=\delta _{jm}\delta _{kn} $ and
as in (\ref{eq:26a}) $\bar{k}=N+1-k $. Then we make use of the fact that
if $V $ is a group element of the corresponding group then
$V^T=SV^{-1}S^{-1} $, i.e.
\begin{equation}\label{eq:ap-2.3}
r_k = V\left\{ \begin{array}{ccccc} 1 & \dots & \hat{\bar{k}} & \dots & N
\\ 1 & \dots & k & \dots & N-1  \end{array} \right\}
\end{equation}
where the `hat` means that the index $\bar{k} $ is missing. Equation
(\ref{eq:ap-2.1}) and (\ref{eq:ap-2.3}) readily give:
\begin{equation}\label{eq:ap-3.1}
r_k r_{\bar{k}} =  {W(1,\dots, \hat{\bar{k}}, \dots,N) \over p_{\bar{k}}
W(1,\dots,N) } \prod_{s=1}^{N} p_s r_s
\end{equation}
Taking the product of (\ref{eq:ap-3.1}) for $k=1 $ to $r $ one is able to
evaluate $\prod_{s=1}^{N} r_k $ in terms of $p_s $ and $\zeta _s $ alone.
Next putting $p_s $, $N $ and $\zeta _s $ as appropriate for the series $
{\bf B}_r $ and ${\bf C}_r $ we derive the expressions for $w_k $
(\ref{eq:w-k-b}), (\ref{eq:r0-b}) and (\ref{eq:w-k-c}).

{}For the ${\bf D}_r $ series the matrix $L(0) $ is of the form:
\begin{equation}\label{eq:ap-3.2}
L(0) = \left( \begin{array}{ccccc|ccccc}
b_1 & a_1 & &  &  & & & & & \\
a_1 & b_2 & &  &  & & & & & \\
& & \ddots & \ddots & \ddots &&&&& \\
 &  & & b_{r-1} & a_{r-1} & a_r & 0 & & & \\
 &  & & a_{r-1} & b_r & 0 & a_r & && \\ \hline
 &  & & a_r & 0 & -b_r & a_{r-1} & && \\
 &  & & 0 & a_r & a_{r-1} & -b_{r-1} & &  &\\
&&&&& \ddots & \ddots & \ddots && \\
&&&&&&& & -b_2 & a_1 \\
&&&&&&& & a_1 & -b_1 \end{array} \right).
\end{equation}
{}For the $k $-th eigenvector we get:
\begin{equation}\label{eq:ap-3.3}
v^{(k)} \simeq \left( \begin{array}{c} r_k p_1 \\ r_k p_2
\zeta _k  \\ \vdots \\
r_k  p_{r-1} \zeta^{r-2} _k \\ r_k p_r (\zeta _k^{r-1} + C/\zeta _k)
\\ r_k p_r (\zeta _k^{r-1} - C/\zeta _k) \\ r_k \zeta _k^{r}
p_{r+2} \\ \vdots \\ r_k \zeta _k^{2r-2} p_{2r}\end{array} \right),
\end{equation}
where $C $ is a coefficient to be calculated below. The symbol
$\simeq $ in (\ref{eq:ap-3.2}) means that in the right hand side we have
omitted terms polynomial in $\zeta _k $ which do not contribute to the
minors of $V $. Note also that $C $ enters into play only when we need a
minor of order $r $ or higher. Such necessity appears in two cases: when
we evaluate the expression for $\tilde{A}_r(t) $ and when we derive the
symmetry relation.

{}From (\ref{eq:eigve}) and the explicit formula for $L(0) $
(\ref{eq:ap-3.2}) we find that $p_k $ in (\ref{eq:ap-3.3}) are given as
follows:
\begin{eqnarray}\label{eq:65.3}
&& p_1=1, \qquad p_k = \prod_{s=1}^{k-1} {1  \over a_s(0) }, \qquad
\mbox{for}\; k=2,\dots, r-1, \nonumber\\
&&\fl p_r = {1  \over 2 } \prod_{s=1}^{r-1} {1  \over a_s(0) }, \qquad
p_{r+1} = {1  \over 2a_r(0) } \prod_{s=1}^{r-2} {1  \over a_s(0) },
\qquad p_{r+2} = {1  \over 2 } \prod_{s=1}^{r} {1  \over a_s(0) },
\nonumber\\
&& p_{r+k} = p_{r+2} \prod_{s=2}^{k-1} {1  \over a_{r-s}(0) }, \qquad
\mbox{for}\; k=3,\dots, r.
\end{eqnarray}
The determinant of $V $ gives:
\begin{equation}\label{eq:65.2}
1=\det V = 2 (-1)^{r-1} C W(1,\dots, 2r) \prod_{s=1}^{2r} {r_s p_s \over
\zeta _s}.
\end{equation}

In analogy with (\ref{eq:ap-2.3}), (\ref{eq:ap-3.1}) we get:
\begin{equation}\label{eq:ap-4.2}
r_p r_{\bar{p}} = {(-1)^{r} 2C\zeta _p \over p_{2r} }
{  W(1,\dots, \hat{\bar{p}}, \dots, 2r) \over W(1, \dots, 2r)}
\prod_{s=1}^{2r} {p_s r_s  \over \zeta _s }.
\end{equation}
Taking again the product $\prod_{p=1}^{r} r_p r_{\bar{p}} $ in
(\ref{eq:ap-4.2}) and substituting the expressions for $p_s $ from
(\ref{eq:65.3}) we find
\begin{equation}\label{eq:ap-5.1}
C=(-1)^{r+1} \prod_{s=1}^{r} \zeta _s,
\end{equation}
and in addition - the relation (\ref{eq:w-k-d}).

Now it is easy to find the expression for the minor of order $r $:
\begin{eqnarray}\label{eq:ap-4.1}
V\left\{ \begin{array}{ccc} 1 &\dots & r\\ i_1 & \dots & i_r \end{array}
\right\} = {1  \over 2 }  \left( 1 + {\zeta _{1} \dots \zeta _{r} \over
\zeta _{i_1} \dots \zeta _{i_r} } \right) W(i_1,\dots,i_r)
\prod_{s=1}^{r}r_{i_s} p_s.
\end{eqnarray}
needed for the derivation of $\tilde{A}_r(t) $ (\ref{eq:Dr-Ar}).

\section{Algebraic details}

The action of $w_0$ on the simple roots is well known
\cite{Bour,GotGro*78}:
\begin{equation}\label{eq:w0}
w_0(\alpha _k)=-\alpha _{\tilde{k}},
\end{equation}
where $\tilde{k}=r-k+1$ for ${\bf A}_r$; $\tilde{k}=k$,  $k=1,\ldots,r$ for
${\bf B}_r$, ${\bf C}_r$ and ${\bf D}_{2n}$. For ${\frak g}\simeq {\bf
D}_{2n+1} $ we have $\tilde{k}=k$ for $k\leq 2n-1$, and $w_0(\alpha
_{2n})=-\alpha _{2n+1} $,  $w_0(\alpha _{2n+1})=-\alpha _{2n} $.
More specifically $w_0 $ acts on the orthonormal basis $\{e_k\} $ in the
root space as follows:
\begin{eqnarray}\label{eq:w0a}
w_0(e_k)= e_{\bar{k}}, \qquad \mbox{for}\; {\bf A}_r, \\
w_0(e_k)=- e_{k}, \qquad \mbox{for}\; {\bf B}_r, {\bf C}_r,
{\bf D}_{2n}, \nonumber
\end{eqnarray}
and for ${\bf D}_{2n+1}  $
\[
w_0(e_k) = -e_k, \qquad \mbox{for}\; k=1,\dots, 2n, \qquad w_0(e_{2n+1}) =
e_{2n+1}.
\]

Next well known fact is that the weight system $\Gamma (\omega ) $ is
determined uniquely by the highest weight $\omega  $. The reconstruction
of the weights $ \gamma \in \Gamma (\omega ) $ is performed by using two
facts: i)~if $\gamma \in \Gamma (\omega ) $ then $w(\gamma) \in \Gamma
(\omega ) $ where $w $ is any element of the Weyl group; besides $\gamma $
and $w(\gamma ) $ have equal multiplicities; ii)~if $\alpha >0 $ is a
positive root and
\begin{equation}\label{eq:ap-1}
{2(\alpha,\omega )  \over (\alpha,\alpha ) } = p>0,
\end{equation}
then $\omega -s\alpha \in \Gamma (\omega ) $ for all $s=1,\dots,p $.

In particular, if $\omega =\omega _k^+ $ and $\alpha =\sum_{s=1}^{r} m_s
\alpha _s $ we see that (\ref{eq:ap-1}) is fulfilled only if $m_k=p>0 $.
Thus we find that generically (i.e. for $k<r $)  along with $\omega_k ^+
$, weights in $\Gamma (\omega ^+_k) $ are also
\[
\gamma _1 =\omega _k^+-\alpha _k, \qquad \gamma _2=\omega _k^+ - (\alpha
_{k-1}+\alpha _k), \qquad \mbox{etc.}
\]

Using the sorting condition (\ref{eq:wech}) we easily find that
\begin{equation}\label{eq:ap-2}
\mathop{\min}\limits_{\gamma \in \Gamma (\omega _k^+) \backslash
\omega _k^+} \left[-(\vec{\kappa},\omega _k^+-\gamma )\right] =
\min \left[\sum_{s=1}^{r} m_s (-\vec{\kappa },\alpha _s)\right] =
-(\vec{\kappa },\alpha _k),
\end{equation}
which proves the estimations in Equations (\ref{eq:Bkas}).

The same method can be applied also when one of the roots satisfies
$(\vec{\kappa },\alpha _m)=0 $. As a consequence of this condition at
least two terms in ${\cal  B}_k(t) $ may have the same asymptotic
behavior.

Here we will first describe the sets of roots $G_p^+(\vec\kappa ) $
see (\ref{eq:139.4}) and then will also outline the proof of
(\ref{eq:139.1}).  Obviously if $(\vec\kappa,\alpha _m)=0 $ only
$G_m^+(\vec\kappa ) $ will be nonempty
and will coincide with $\{\alpha _m\} $ and therefore $\Gamma
_{p,+}(\omega _p^+)\backslash \{\omega _p^+
\}$ while $\Gamma _{m,+}(\omega _m^+)\backslash \{\omega _m^+,\omega
_m^+-\alpha
\}$ where $\alpha =\alpha _m +\sum_{s}^{}m_s \alpha _s $.
The minimum of $-2(\vec\kappa,\omega _m^+-\alpha ) $ will be achieved if
we limit ourselves with roots $\alpha  $ of height 2. Now it remains to
take into account that $\alpha _m+\alpha _s $ is a root if and only if
$(\alpha _m,\alpha _s)< 0 $. The corresponding result  for $t\to -\infty
$ is obtained by acting with $w_0 $.  This proves (\ref{eq:139.1}).

We finish this appendix by describing the sets of roots $G_p^+(\vec\kappa
) $ for $(\vec\kappa,\alpha _m)=(\vec\kappa,\alpha _m)=0 $. First, if
$(\alpha _m,\alpha _p)=0 $ then $G_m^+(\vec\kappa )=\{\alpha _m\} $,
$G_p^+(\vec\kappa )=\{\alpha _p\} $ and all the others $G_k^+(\vec\kappa
)=\{\emptyset \} $. If however $(\alpha _m,\alpha _p)<0 $ the situation
becomes more interesting. In the generic case $(\alpha _m,\alpha _p)=-1 $
we find
\[
G^+_m(\vec\kappa )=\{\alpha _m,\alpha _m+\alpha _p\}, \qquad
G_p^+(\vec\kappa )=\{\alpha _p,\alpha _m+\alpha _p\}.
\]
The only two exceptions of this rule for the classical series are $m=r-1
$, $p=r $ for ${\frak  g}\simeq {\bf B}_r $ and ${\bf C}_r $. Then we
have:
\[ \fl
G^+_{r-1}(\vec\kappa )=\{\alpha _{r-1},\alpha _{r-1}+\alpha _r, \alpha
_{r-1}+2\alpha _r \}, \qquad
G^+_{r}(\vec\kappa )=\{\alpha _{r},\alpha _{r-1}+\alpha _r, \alpha
_{r-1}+2\alpha _r \},
\]
for ${\frak  g}\simeq {\bf B}_r $  and
\[ \fl
G^+_{r-1}(\vec\kappa )=\{\alpha _{r-1},\alpha _{r-1}+\alpha _r, 2\alpha
_{r-1}+\alpha _r \}, \qquad
G^+_{r}(\vec\kappa )=\{\alpha _{r},\alpha _{r-1}+\alpha _r, 2\alpha
_{r-1}+\alpha _r \},
\]
for ${\frak  g}\simeq {\bf C}_r $. These last relations allow us to
calculate the asymptotics of ${\cal B}^\pm_{k,\rm as} $ for all possible
values of $k $ for $I_{\rm bs}=\{m,p\} $. Then it is not difficult to
insert them in (\ref{eq:5.2}) and evaluate the asymptotic behavior of all
$q_k(t) $. Several examples of such calculations were presented in
Section~4 above.

\end{document}